\documentclass[12pt]{article}
\pdfoutput=1
\usepackage{cite}
\usepackage{amssymb}
\usepackage{amsmath}
\usepackage{bm} 
\usepackage{graphicx}
\usepackage{color}
\usepackage{units}
\usepackage{xcolor}
\usepackage{dsfont}
\usepackage{enumerate}
\usepackage{hyperref}
\usepackage{setspace}
\usepackage[latin1]{inputenc}
\definecolor{nicered}{rgb}{0.7,0.1,0.1}
\definecolor{nicegreen}{rgb}{0.1,0.5,0.1}
\hypersetup{colorlinks,citecolor= nicegreen,linkcolor= nicered}

\def\L{{\cal L}}

\newcommand{\be}  {\begin{equation}}
\newcommand{\ee}  {\end{equation}}

\def\e6{E(6)}
\def\10{SO(10)}
\def\21{SA(2) $\otimes$ U(1) }
\def\321{$\mathrm{SU(3) \otimes SU(2) \otimes U(1)}$ }

\def\422{SA(4) $\otimes$ SA(2) $\otimes$ SA(2)}
\def\roughly#1{\mathrel{\raise.3ex\hbox{$#1$\kern-.75em
      \lower1ex\hbox{$\sim$}}}} \def\lsim{\roughly<}
\def\gsim{\roughly>}

\def\lsim{\raise0.3ex\hbox{$\;<$\kern-0.75em\raise-1.1ex\hbox{$\sim\;$}}}
\def\gsim{\raise0.3ex\hbox{$\;>$\kern-0.75em\raise-1.1ex\hbox{$\sim\;$}}}

\def \epsilon {\varepsilon} 
\newcommand{\hc}{\text{h.c.}}
\newcommand{\Br}{\text{Br}}
\newcommand{\matrixx}[1]{\begin{pmatrix} #1 \end{pmatrix}} 

\nopagebreak
\allowdisplaybreaks
\begin{document}
\begin{titlepage}
 \begin{flushright}
ULB-TH/16-12
 \end{flushright}

  \newcommand{\AddrIPM}{{\sl \small $^a$ School of Physics, Institute for
      Research in Fundamental Sciences (IPM),\\ \sl \small
      P.O.~Box 19395-5531, Tehran, Iran}}

  \vspace*{0.5cm}

  \newcommand{\AddrBLU}{{\sl \small $^b$ Service de Physique Th\'eorique, Universit\'e  Libre de Bruxelles,\\ \sl \small
      Boulevard du Triomphe, CP225, 1050 Brussels, Belgium}}
  \vspace*{0.5cm}

\begin{center}
  \textbf{\large Neutrinophilic Non-Standard Interactions }

  \vspace{1cm}

  Yasaman Farzan\footnote{E-mail address: {\tt yasaman@theory.ipm.ac.ir}}$^a$  and Julian Heeck\footnote{E-mail address: {\tt Julian.Heeck@ulb.ac.be}}$^b$
  \vspace*{0.4cm}\\
  \AddrIPM \\
  \AddrBLU
\end{center}
\vspace*{0.2cm}

\begin{abstract}
 \onehalfspacing

We construct UV-complete models for non-standard neutrino interactions mediated by a sub-GeV gauge boson $Z'$ coupled to baryon number $B$ or $B-L$. A flavor-dependent $Z'$ coupling to neutrinos is induced by mixing a $U(1)'$-charged Dirac fermion with the active neutrinos, naturally suppressing  flavor violation or non-universality of the charged leptons to the loop level.
We show that these models can give rise to large flavor-conserving as well as flavor-violating non-standard neutral-current neutrino interactions potentially observable in current or future oscillation experiments such as DUNE without being in conflict with other constraints such as neutrino scattering or lepton-flavor-violating decays. In particular, the LMA-Dark solution to the solar-neutrino anomaly can be obtained  for $U(1)_B$, but not for $U(1)_{B-L}$.

\end{abstract}
\end{titlepage}

\setcounter{footnote}{0}
\section{Introduction}
\label{intro}

The three-neutrino  mass and mixing scheme has been triumphant in explaining solar, atmospheric, reactor, and long-baseline neutrino data. Thanks to the extensive running and upcoming neutrino experimental programs such as T2K, NO$\nu$A, and DUNE we are entering the neutrino precision era. Especially the DUNE and T2HK experiments are expected to make it  possible to measure the yet unknown neutrino parameters such as the Dirac CP-violating  phase $\delta_\mathrm{CP}$, the octant of the atmospheric mixing angle $\theta_{23}$, and the neutrino mass ordering (normal vs.~inverted). However, these claims are valid only under the assumption of the standard three-neutrino paradigm with standard interactions between neutrinos and matter fields. If the neutral-current interactions of neutrinos with matter fields deviate from the Standard Model (SM), the neutrino propagation in matter between the source and detector at long-baseline experiments is affected.
Such Non-Standard neutrino Interactions (NSI) can be parametrized by the effective Lagrangian
\begin{align}
\L_\text{NSI} = -2\sqrt2 G_F \,\epsilon_{\alpha\beta}^{f X}\, (\bar\nu_\alpha \gamma^\mu P_L \nu_\beta)(\bar f \gamma_\mu P_X f)\,,
\label{eq:NSI}
\end{align}
where $P_{R/L} \equiv (1\pm \gamma_5)/2$ is the chirality projection operator, $f\in\{e,u,d\}$ specifies the matter particles, and $\alpha, \beta\in \{e,\mu,\tau\}$ denote the neutrino flavor.
The dimensionless coefficients~$\epsilon_{\alpha\beta}^{f X}$ have been normalized to the electroweak strength, $2\sqrt2 G_F\simeq (\unit[174]{GeV})^{-2}$.
Only the vector coupling is relevant for neutrino oscillations, so we define $\epsilon_{\alpha\beta}^f \equiv \epsilon_{\alpha\beta}^{f L}+\epsilon_{\alpha\beta}^{f R}$ as the quantity of interest in the following.
 As has recently been shown in a series of papers \cite{Friedland:2012tq,Masud:2015xva,deGouvea:2015ndi,Coloma:2015kiu,Masud:2016bvp,Masud:2016gcl,Blennow:2016etl,Bakhti:2016gic}, if neutrino interactions with matter fields ($e$, $u$ and $d$) deviate from those in the SM, new degeneracies appear making an unambiguous derivation of the unknown neutrino parameters impossible. In particular, as shown in Ref.~\cite{Forero:2016cmb}, neutral-current NSI can mimic the effect of $\delta_\mathrm{CP}$ at DUNE even if all the sources of CP violation in the leptonic sector (both standard Dirac phases and phases of the new couplings) vanish.
 Moreover, the determination of the octant of $\theta_{23}$ can become problematic in the presence of complex $\epsilon_{e \tau}$ or $\epsilon_{e \mu}$~\cite{Agarwalla:2016fkh}. As has been shown in Ref.~\cite{Bakhti:2016prn}, combining the results of very-long-baseline experiments like NO$\nu$A with the proposed medium-baseline ($L\sim \unit[150]{km}$) experiment MOMENT can help to solve this degeneracy.

It is remarkable that in addition to the standard LMA (Large Mixing Angle) solution to the solar-neutrino anomaly with $\theta_{12}<\pi/4$ and  $\epsilon_{\alpha \beta}^f=0$, there is another solution (called LMA-Dark solution) with $\theta_{12}>\pi/4$ and $\epsilon_{\mu \mu}^{u,d}-\epsilon_{ee}^{u,d}\simeq \epsilon_{\tau \tau}^{u,d}-\epsilon_{ee}^{u,d}\sim 1$~\cite{Gonzalez-Garcia:2013usa,Miranda:2004nb,Escrihuela:2009up}. We discuss this solution later on. 
Using instead the standard LMA solution, one can derive the current $90 \%$~C.L.~bounds on the values of $\epsilon^u_{\alpha \beta}$ from global oscillation data~\cite{Gonzalez-Garcia:2013usa}. Taking the conservative values from Ref.~\cite{Coloma:2015kiu}, these read
\begin{align}
|\epsilon_{e\mu}^u+\epsilon_{e\mu}^d| <0.12 \,, \qquad |\epsilon_{e\tau}^{u}+\epsilon_{e\tau}^d| <0.18 \,, \qquad |\epsilon_{\mu\tau}^{u}+\epsilon_{\mu\tau}^d| <0.018 \,,   \\
0.11<\epsilon_{ee}^u+\epsilon_{ee}^d-\epsilon_{\tau\tau}^u-\epsilon_{\tau\tau}^d<0.60 \,,  \quad {\rm and} \quad -0.04<\epsilon_{\mu\mu}^u+\epsilon_{\mu\mu}^d-\epsilon_{\tau\tau}^u-\epsilon_{\tau\tau}^d<0.037 \,,
\label{currentBounds}
\end{align}
assuming $\epsilon^{e}=0$.
Remember that the neutrino-oscillation pattern does not change if we replace the Hamiltonian $\mathcal{H}$ governing neutrino evolution in time with $\mathcal{H}- {1\!\!1} a$, where ${1\!\!1}$ is the identity matrix in flavor space and $a$ is an arbitrary number. As a result, neutrino-oscillation data can only provide information on the \emph{splitting} of the diagonal elements. Using the priors of Ref.~\cite{Gonzalez-Garcia:2013usa} (with a best fit deviating from zero), it has been shown in Ref.~\cite{Coloma:2015kiu} that T2HK together with DUNE can improve these bounds down to
\begin{align}
|\epsilon_{e\mu}^u+\epsilon_{e\mu}^d| <0.024 \,, \qquad |\epsilon_{e\tau}^{u}+\epsilon_{e\tau}^d| <0.08 \,, \qquad |\epsilon_{\mu\tau}^{u}+\epsilon_{\mu\tau}^d| <0.012 \,,   \\
0.017<\epsilon_{ee}^u+\epsilon_{ee}^d-\epsilon_{\tau\tau}^u-\epsilon_{\tau\tau}^d<0.43 \,,  \quad {\rm and} \quad -0.027<\epsilon_{\mu\mu}^u+\epsilon_{\mu\mu}^d-\epsilon_{\tau\tau}^u-\epsilon_{\tau\tau}^d<0.025 \,.
\label{futureBounds}
\end{align}

 From a theoretical point of view, the question arises whether it is possible to build a consistent renormalizable model that gives rise to an effective Lagrangian of the form of Eq.~\eqref{eq:NSI} with large enough $\epsilon$ to be observable in neutrino experiments ({\it i.e.,} $|\epsilon|\gtrsim 0.05$). The first solution which comes to mind is introducing a heavy intermediate state $X$ with coupling to matter fields and neutrinos which has so far escaped direct production because of its large mass $M_X$.
 Integrating out this heavy state can easily give rise to the four-fermion interactions of Eq.~(\ref{eq:NSI}) but the value of $\epsilon$ is suppressed by $M_W^2/M_X^2\ll 1$. An alternative approach which has been incorporated by Refs.~\cite{Farzan:2015doa,Farzan:2015hkd,Machado:2016fwi} is to introduce a new $U(1)^\prime$ gauge interaction with a relatively light gauge boson~$Z'$, with mass $M_{Z'}\sim$few~10~MeV.\footnote{Taking the $Z'$ much lighter leads to long-range interactions with different phenomenology~\cite{Joshipura:2003jh,Grifols:2003gy,Bandyopadhyay:2006uh,GonzalezGarcia:2006vp,Samanta:2010zh,Heeck:2010pg,Davoudiasl:2011sz,Chatterjee:2015gta}.}
In this class of models, the new gauge boson has so far escaped detection because of the smallness of its coupling rather than its large mass. Matter effects on propagation of neutrinos are induced by $t$-channel forward scattering of neutrinos ({\it i.e.,} scattering with zero energy--momentum transfer); as a result, we can still use the effective Lagrangian in Eq.~(\ref{eq:NSI}) even if the mass $M_{Z'}$ of the intermediate $Z^\prime$ boson is much smaller than the typical energies of the neutrinos propagating in the medium. However, for neutrino scattering experiments such as CHARM or NuTeV with an energy--momentum transfer $q$ much larger than the $Z^\prime$ mass ({\it i.e.,} $q^2 \gg M_{Z'}^2$), we can no longer invoke the effective Lagrangian formalism. At these scattering experiments, the ratio of the amplitude of the new contribution to the SM amplitude  is  therefore suppressed by $\epsilon M_{Z'}^2/q^2$  and is below the sensitivity limit~\cite{Farzan:2015doa}.

An $SU(2)_L\times U(1)_Y$ invariant realization of Eq.~(\ref{eq:NSI}) typically implies that \emph{charged} leptons should have similar new interactions as neutrinos. Since the bounds on such new interactions of charged leptons (especially on $e$ and $\mu$) are strong, model building is far from trivial. The challenge is even more severe if we want to build a model which gives rise to lepton-flavor-violating (LFV) NSI ({\it i.e.,} $\epsilon_{\alpha \beta}|_{\alpha \ne \beta} \ne 0$) because of very strong bounds from associated charged-lepton flavor violating (CLFV) processes such as $\ell_\alpha \to \gamma \ell_\beta$, $\ell_\alpha \to \ell_\beta\ell_\gamma\ell_\delta$, and $\ell_\alpha \to Z^\prime \ell_\beta$~\cite{Heeck:2016xkh}.

 In this article, we present models based on  new $U(1)^\prime$ gauge symmetries with a light gauge boson $Z^\prime$ which can give rise to both lepton-flavor conserving and LFV neutral-current NSI without inducing similar couplings to the charged leptons. This is done by introducing a new Dirac fermion $\Psi$ charged under $U(1)^\prime$ which is mixed  with  neutrinos by  Yukawa couplings to a new scalar doublet $H^\prime$. We are interested in a form of NSI that affects neutrino propagation in matter but not neutrino interaction at source and detector.
Furthermore, our NSI are always vector-like, i.e.~fulfil $\epsilon_{\alpha \beta}^{fL}=\epsilon_{\alpha \beta}^{fR}$.
This is convenient, because otherwise the axial part of the current changes the cross section of Deuteron dissociation at SNO ({\it i.e.,} $D+\nu \to p+n+\nu$). This process is not influenced when $\epsilon_{\alpha \beta}^{qL}=\epsilon_{\alpha \beta}^{qR}$, so the consistency of the total neutral current rate at SNO with the total neutrino flux predicted by the standard solar model is maintained despite large $\epsilon_{\alpha \beta}^{uL}=\epsilon_{\alpha \beta}^{uR}$ and $\epsilon_{\alpha \beta}^{dL}=\epsilon_{\alpha \beta}^{dR}$~\cite{Miranda:2004nb,Davidson:2003ha}.

The rest of this paper is organized as follows: In Sec.~\ref{model}, we present a class of models that can give rise to large NSI. We discuss the constraints from the charged LFV bounds and scattering of solar neutrinos at the solar neutrino experiments as well as at dark matter direct-detection experiments. We also discuss possible routes to UV-complete the model. In Sec.~\ref{implications}, we discuss the observational consequences of the model. Results are summarized in Sec.~\ref{summary}.

\section{Neutrinophilic LFV}
\label{model}

In this section, we describe our model which is based on a $U(1)^\prime$ gauge symmetry with a light, MeV--GeV,  gauge boson $Z^\prime$. In order to avoid tree-level flavor-changing neutral currents, we assume a universal $Z'$ coupling to baryons $g_B$ -- so quarks carry $g_B/3$ -- and a universal lepton coupling $g_\ell$ (including to right-handed neutrinos $\nu_R$).
Moreover, the SM scalar doublet $H$ is assumed to be neutral under this $U(1)'$, so all fermions acquire Dirac masses after electroweak symmetry breaking. In models with $g_\ell\ne 0$, an additional singlet scalar $S_R$ with $U(1)^\prime$ charge equal to $ -2 g_\ell$ is required to generate a Majorana mass for the right-handed neutrinos $S_R  \bar \nu^c_R \nu_R$, which gives a seesaw mass $\mathcal{M}_\nu \propto \langle H\rangle^2/\langle S_R\rangle$ for the light neutrinos. In models with $g_\ell=0$, neutrinos obtain mass via canonical seesaw without any need to introduce a new scalar. To generate a neutrino-flavor-dependent $Z'$ coupling we introduce a Dirac fermion $\Psi$ with mass $M_\Psi$ and $U(1)'$ charge $g_\Psi$ as well as a second scalar doublet $H'$ with charge $g_\Psi-g_\ell$ (otherwise the same quantum numbers as $H$), which allows for the Yukawa couplings
\begin{align}
\L = -\sum_\alpha y_\alpha \overline{L}_\alpha \tilde H' P_R \Psi + \hc ,
\label{eq:newyukawa}
\end{align}
where $\tilde H'\equiv i \sigma_2 (H')^*$.
The light neutrinos and $\Psi$ share a mass matrix (in compact form)
\begin{align}
\L = \frac12 (\bar\nu^c, \bar\Psi_L^c, \bar\Psi_R) \matrixx{\mathcal{M}_\nu & 0 & y\langle H'\rangle\\ 0 & 0 & M_\Psi\\ y\langle H'\rangle & M_\Psi & 0 } \matrixx{\nu\\\Psi_L\\\Psi_R^c} +\hc ,
\end{align}
which leads to mixing among $\Psi_L$ and $\nu$. In the limit $\mathcal{M}_\nu, y\langle H'\rangle \ll M_\Psi$, the mixing angles are small and can be written as~\cite{Fox:2011qd}
\begin{align}
\label{k-a}
\kappa_\alpha = \frac{y_\alpha\langle H'\rangle}{M_\Psi} = \frac{y_\alpha v \cos\beta}{\sqrt2 M_\Psi} \,,
\end{align}
which can in general be complex. 
Note that despite the mixing with $\Psi$, the active neutrinos remain massless in the limit $\mathcal{M}_\nu = 0$.
As in standard two-Higgs-doublet model (2HDM) notation~\cite{Branco:2011iw}, we have defined an angle $\beta$ via $\tan\beta \equiv\langle H\rangle/\langle H'\rangle$, and $v\simeq\unit[246]{GeV}$. The contribution of $\langle H'\rangle$ to $M_{Z'}^2$ can be written as $(g_\Psi-g_\ell)^2 v^2 \cos^2 \beta$, which should be summed with the contributions from the vacuum expectation values (VEVs) of other scalars charged under $U(1)^\prime$. Since we want the $Z^\prime$ to be light, we demand
\begin{align}
\label{BoundsOnbeta}
\cos \beta \leq  4\times 10^{-4} \left( \frac{M_{Z'}}{\unit[10]{MeV}}\right) \left(\frac{0.1}{|g_\Psi-g_\ell|} \right).
\end{align}
The relevant $Z'$ interactions $Z'_\mu \left( g_\ell \bar\nu\gamma^\mu P_L \nu + g_\Psi \bar\Psi \gamma^\mu \Psi\right)$
can be rewritten in the mass basis as
\begin{align}
 Z'_\mu  \left[\sum_{\alpha,\beta}(g_\ell \delta_{\alpha\beta} + g_\Psi \kappa_\alpha^* \kappa_\beta) \bar\nu_\alpha\gamma^\mu P_L \nu_\beta + \sum_\alpha (g_\ell - g_\Psi)  \left[ \kappa_\alpha^* \bar\nu_\alpha\gamma^\mu P_L \Psi + \kappa_\alpha \bar\Psi \gamma^\mu P_L \nu_\alpha \right]+ g_\Psi \bar\Psi \gamma^\mu \Psi\right] ,
\label{eq:Zprimecouplings}
\end{align}
where the mass eigenstate $\Psi$ is approximately the same Dirac fermion as above and we neglected terms of order $g_\ell \kappa^*\kappa$.
The crucial results are the off-diagonal and non-universal $Z'$ couplings to the light neutrinos via $g_\Psi \kappa_\alpha^* \kappa_\beta$, while the charged-lepton $Z'$ couplings remain diagonal. (Rotating the light neutrinos to their mass eigenstates merely redefines the $\kappa$.)
We then obtain our desired NSI coefficients
\begin{align}
\epsilon_{\alpha\beta}^{u}=\epsilon_{\alpha\beta}^{d} \simeq  \frac{g_B g_\Psi \kappa_\alpha^* \kappa_\beta}{6\sqrt2 G_F M_{Z'}^2} \,, &&
\epsilon_{\alpha\beta}^{e} \simeq  \frac{g_\ell g_\Psi \kappa_\alpha^* \kappa_\beta}{2\sqrt2 G_F M_{Z'}^2} \,.
\label{eq:ourNSI}
\end{align}
Note that the NSI coefficients cannot be chosen completely arbitrary, as $\kappa^* \kappa^T$ is only a Hermitian rank-1 matrix with three parameters. In particular, $|\epsilon_{\alpha \beta}|= \sqrt{\epsilon_{\alpha \alpha} \epsilon_{\beta \beta}}$. Introducing more copies of $\Psi$ allows for more freedom, as it replaces $g_\Psi \kappa^* \kappa^T \to \sum_j g_{\Psi_j} \kappa_{\Psi_j}^* \kappa_{\Psi_j}^T $. However, due to the Cauchy--Schwarz inequality we still have $ |\epsilon_{\alpha \beta}|\leq \sqrt{\epsilon_{\alpha \alpha} \epsilon_{\beta \beta}}$.

The limit of interest is $\kappa \ll 1$ in order to suppress deviations of $U_\text{PMNS}$ from unitarity~\cite{Antusch:2014woa} and CLFV via Eq.~\eqref{eq:newyukawa} (we come back to this issue later). To still generate large NSI we then need a rather light $Z'$. The next section is devoted to a survey of possible $U(1)'$ generators.

Notice that since $\Psi_R$ and $\Psi_L$ have the same $U(1)^\prime$ charge, they do not induce any anomaly so the value for $g_\Psi$ is independent of $g_{\ell,B}$. $\Psi$ cannot be lighter than a few MeV, otherwise it contributes as an extra relativistic degree of freedom in the early Universe. On the other hand, it cannot be heavier than a few GeV because taking $y$ in the perturbative region, $ \cos \beta $ below the bound in Eq.~\eqref{BoundsOnbeta} and $\kappa_\alpha \sim 0.03$ (leading to sizeable $\epsilon$ while still satisfying the unitarity bounds on $U_\text{PMNS}$), from Eq.~\eqref{k-a} we find
\begin{align}
\label{mPsiBound}
M_\Psi <{\rm few~GeV} \left(\frac{M_{Z'}}{\unit[10]{MeV}}\right)\left( \frac{0.2}{g_\Psi}\right) \left(\frac{0.03}{\kappa}\right).
\end{align}
The actual right-handed neutrinos $\nu_R$ that give rise to the light neutrino masses are assumed to be sufficiently heavy and weakly mixed such that they can be ignored in the following.

\subsection{\texorpdfstring{$U(1)'$}{U(1)'} groups of interest\label{group}}

To have observable effects on neutrino propagation in matter,
$Z'$-mediated NSI require  couplings to neutrinos and to matter particles, i.e.~electrons or first-generation quarks. We have shown how to couple the $Z'$ to neutrinos by mixing the neutrinos with a $U(1)'$-charged Dirac fermion; see Eq.~\eqref{eq:Zprimecouplings}. We are left with the task to couple the $Z'$ to matter. As stated above, flavor-changing neutral currents are most easily avoided by generation-independent couplings, so we restrict ourselves to the baryon and lepton number symmetries.\footnote{Even without a direct $Z'$ coupling to SM fermions we inherit a $Z'$ coupling to the hypercharge current, courtesy of kinetic mixing~\cite{Galison:1983pa,Holdom:1985ag}. For $Z'$ masses below the electroweak scale, this is equivalent to a coupling to electric charge, which does not induce NSI in \emph{neutral} matter. For current limits, see e.g.~Ref.~\cite{Batley:2015lha}.}
Gauging classically conserved charges such as baryon number $B$~\cite{Lee:1955vk,Pais:1973mi}, lepton number $L$~\cite{Okun:1969ey}, and $B-L$~\cite{Carlson:1986cu} has been extensively discussed in the literature. Let us for now ignore the newly introduced particles of the last section (or set $g_\Psi = 0$) and study the $Z'$ parameter space for the SM-fermion couplings.

\begin{figure}[t]
\includegraphics[width=0.48\textwidth]{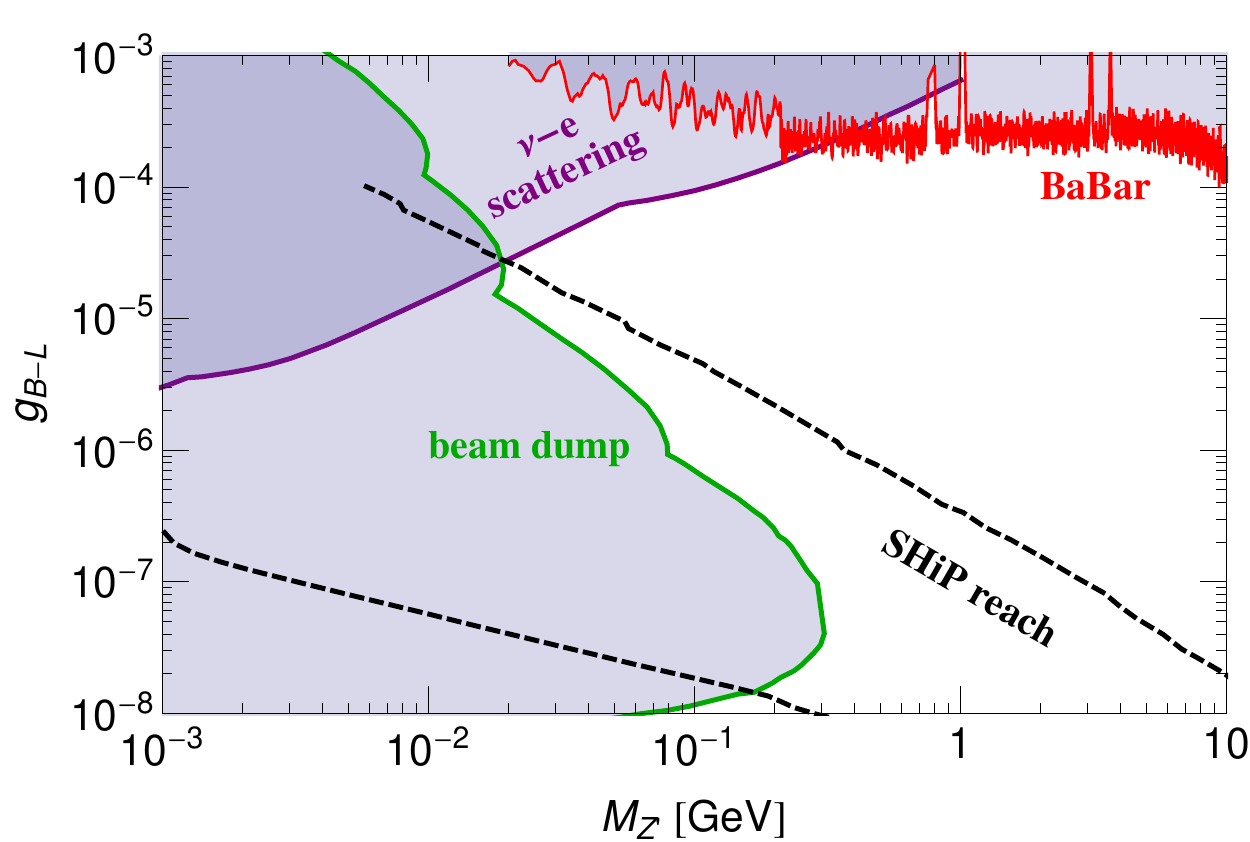}
\includegraphics[width=0.48\textwidth]{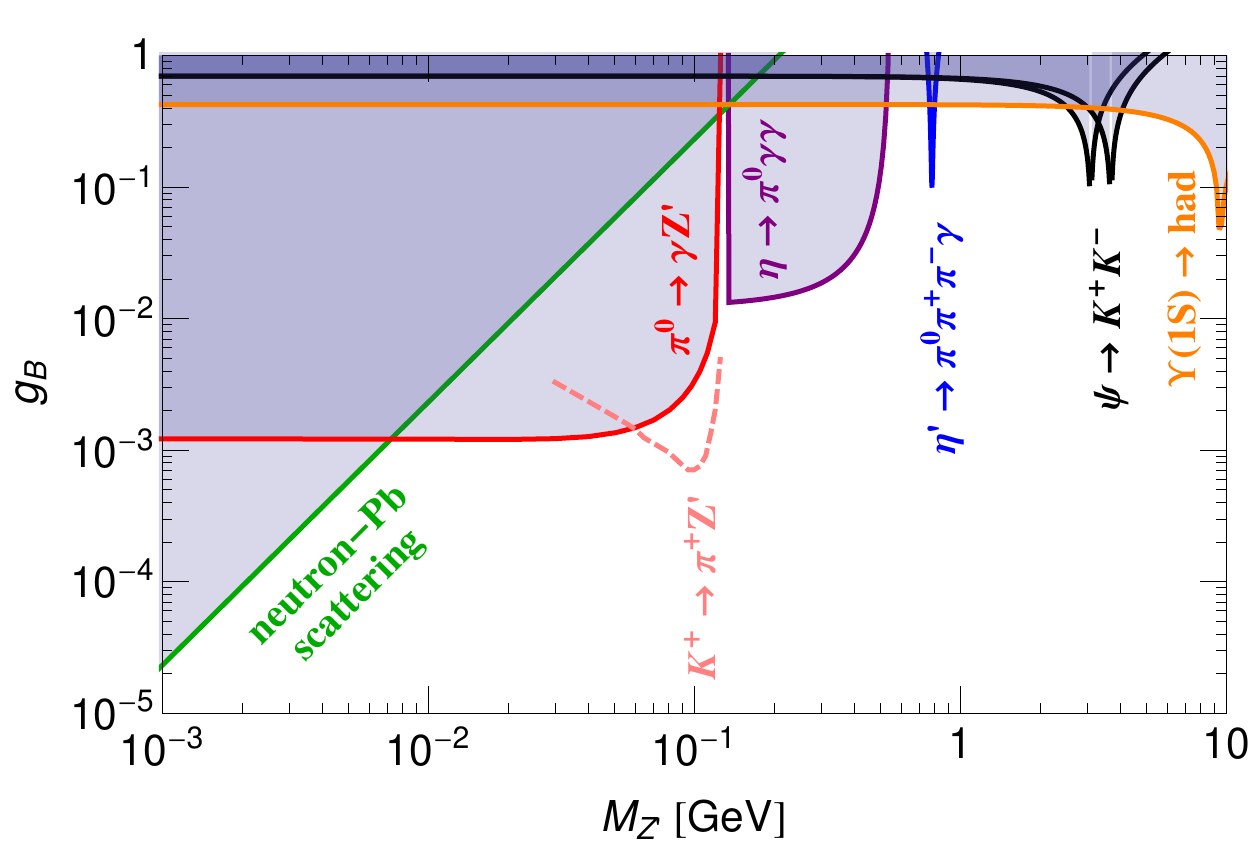}
\caption{
Parameter space of a gauge boson $Z'$ coupled to $B-L$ (left) or $B$ (right). The shaded areas are excluded at $90\%$~C.L.
A $Z'$ coupled to $L$ essentially yields the $B-L$ constraints.
}
\label{fig:limits}
\end{figure}

For $U(1)_{B-L}$, we follow Ref.~\cite{Heeck:2014zfa} to translate the limits from beam-dump experiments~\cite{Andreas:2012mt}, BaBar~\cite{Lees:2014xha}, and $\nu_{e,\mu}$--$e^-$ scattering data~\cite{Harnik:2012ni,Bilmis:2015lja} (see Fig.~\ref{fig:limits} (left)) -- assuming $ M_{Z'} < 2 M_{\nu_R}$.
Also shown is the potential reach of the proposed SHiP experiment~\cite{Alekhin:2015byh}, adopted from Refs.~\cite{Gorbunov:2014wqa,Kaneta:2016vkq}.
The limits for a $U(1)_L$ gauge boson give slightly stronger bounds for the region $M_{Z'}\gtrsim \unit{GeV}$ due to the absence of hadronic decay channels and hence larger leptonic branching ratios.
For $U(1)_B$, the limits on the gauge coupling $g_B$ are much weaker (see Fig.~\ref{fig:limits} (right)). The relevant sub-GeV $Z'$ production and decay branching ratios are given in Ref.~\cite{Tulin:2014tya}; most importantly, the mode $Z'\to \pi\pi$ is suppressed, making $Z'\to\pi^0\gamma$ ($\pi^+\pi^-\pi^0$) dominant for $M_{Z'}\lesssim \unit[0.6]{GeV}$ ($M_{Z'}\gtrsim \unit[0.6]{GeV}$).
Limits come from ${}^{208}$Pb--neutron scattering~\cite{Barbieri:1975xy,Leeb:1992qf,Barger:2010aj}, pion decay $\pi^0\to \gamma Z'$~\cite{Gninenko:1998pm,Altegoer:1998qta}, $\eta\to \gamma Z'\to\gamma\gamma\pi^0$ and $\eta'\to \gamma Z'\to \gamma \pi^+\pi^-\pi^0$ decays~\cite{Tulin:2014tya}, $J/\Psi, \Psi (2S) \to K^+K^-$~\cite{Dobrescu:2014fca}, and hadronic $\Upsilon (1S)$ decays~\cite{Carone:1994aa,Aranda:1998fr}. For $M_{Z'}<m_\pi$ the $Z'$ is essentially stable and invisible, so we can adopt the limit from $K^+\to\pi^+\nu\bar\nu$~\cite{Artamonov:2009sz} derived in Ref.~\cite{Batell:2014yra} (we show the most optimistic limit, i.e.~with cutoff $\Lambda_\text{IR}=m_\rho$).
Some of these limits come with additional uncertainties that make the assignment of a confidence level difficult; we refer the reader to the original articles for details.
The hadronic-decay limits could be improved and refined with Breit--Wigner-peak searches of the $Z'$ final states $\pi^0 \gamma$ and $\pi^+\pi^-\pi^0$~\cite{Tulin:2014tya}.
Additional bounds can be derived from astrophysics and cosmology.

We have neglected kinetic mixing~\cite{Galison:1983pa,Holdom:1985ag} so far, even though it is technically unavoidable. This makes a huge difference in particular for $U(1)_B$, because as long as we set $g_\Psi=0$, it implies $\text{Br} (Z'\to e^+e^-)\simeq 1$ for $2m_e <M_{Z'} < m_\pi$, assuming the loop-induced $Z'\to 3\gamma$ is sufficiently suppressed~\cite{Nelson:1989fx}. This reintroduces e.g.~beam-dump limits on the kinetic mixing angle $\xi$~\cite{Essig:2013lka}. If $\xi$ is large enough so that the $Z'$ boson with $M_{Z'} <m_\pi$ decays promptly to electrons, the $\pi^0\to\gamma Z'$ limit is replaced by the NA48/2 limit from Ref.~\cite{Batley:2015lha}, which is of similar order.
An interesting limiting case was pointed out in Ref.~\cite{Heeck:2014zfa}: if the kinetic mixing angle is opposite in sign but of similar magnitude as the $B-L$ coupling, we obtain a $Z'$ coupling dominantly to the \emph{neutral} fermions, neutrons and neutrinos (with $g_n \simeq -g_\nu$); the couplings $g_p \simeq - g_e$ can be highly suppressed. This severely loosens most of the strong constraints of Fig.~\ref{fig:limits}, including the pion-decay bounds~\cite{Feng:2016jff}, except for the limit from neutron--Pb scattering. A similar limiting case can be considered for $U(1)_B$, where kinetic mixing could cancel the coupling to protons, leaving $Z'$ couplings to neutrons and electrons.
We continue to ignore the kinetic mixing angle, but it should be kept in mind that this additional parameter could either strengthen or weaken the bounds of Fig.~\ref{fig:limits}, without affecting the NSI parameters $\epsilon$ (because we are interested in neutrino propagation through \emph{electrically neutral} matter).

\begin{figure}[t]
\centering
\includegraphics[width=0.48\textwidth]{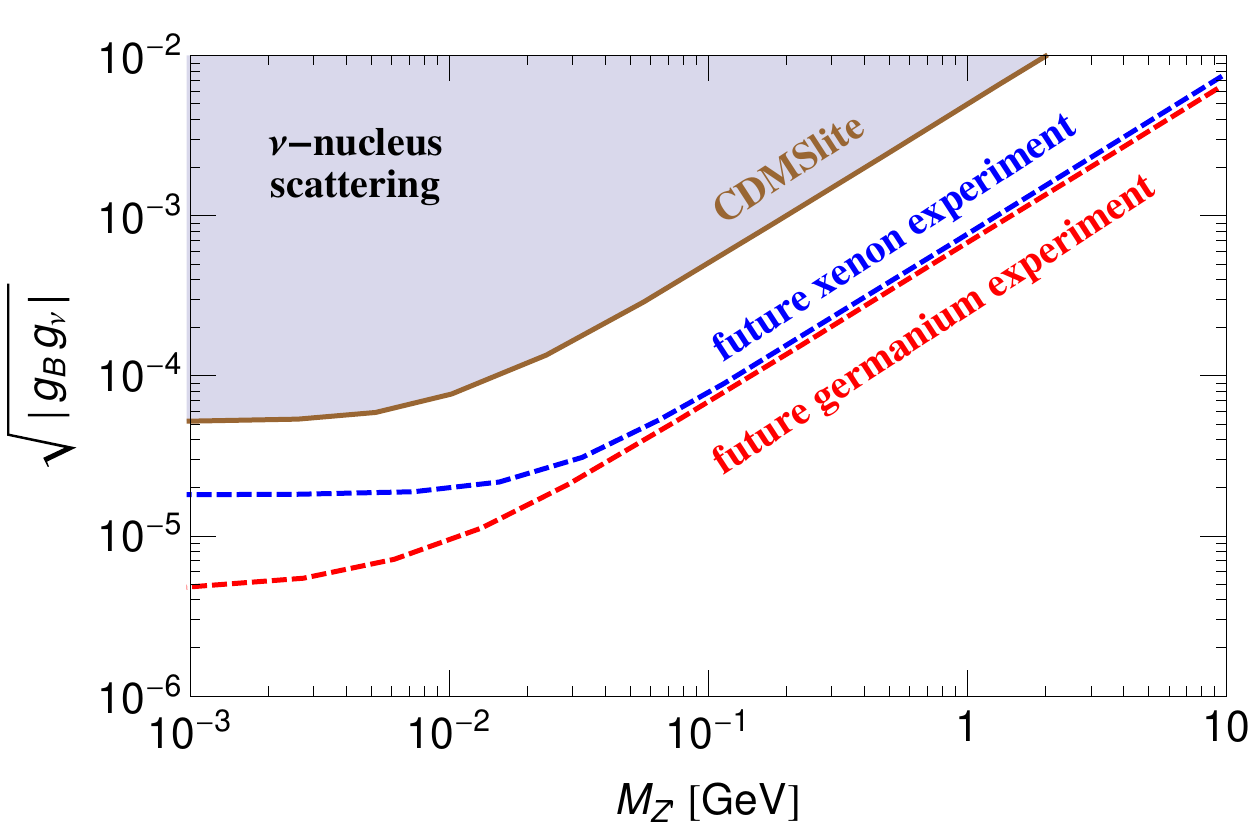}
\caption{
Approximate $90\%$~C.L.~bounds on $\sqrt{|g_B g_\nu|}$ from solar-neutrino nuclear recoils in CDMSlite and optimistic projections for second-generation xenon (e.g.~LUX--ZEPLIN) and germanium experiments (e.g.~SuperCDMS SNOLAB), adapted from Ref.~\cite{Cerdeno:2016sfi}.
}
\label{fig:neutrino_nucleon_scattering}
\end{figure}

The additional $Z'$ couplings to neutrinos $g_\nu = g_\Psi \kappa_\alpha^* \kappa_\beta$ that arise for $g_\Psi \neq 0$ even in the $U(1)_B$ case of course lead to new bounds on top of those described so far. For the $B-L$ case this merely rescales the existing bounds, due to the larger invisible decay rate $Z'\to \bar\nu\nu$ which dilutes the beam-dump limits and the potentially stronger $\nu$--$e$ scattering.\footnote{We ignore the possibility that $g_\nu$ has the opposite sign of $g_{B-L}$ and could thus \emph{soften} scattering constraints, at least for some flavors.}
Qualitatively new bounds emerge for $U(1)_B$ from neutrino--nucleon scattering, proportional to $g_\nu g_B$. A recent study~\cite{Cerdeno:2016sfi} of solar-neutrino scattering rates in dark matter direct detection experiments provides approximate bounds and future projections, shown in Fig.~\ref{fig:neutrino_nucleon_scattering}. Here we ignore the flavor composition of solar neutrinos and simply treat the $Z'$ couplings as diagonal and flavor universal.
Note that we cannot use the effective NSI Lagrangian from Eq.~\eqref{eq:NSI} to describe this scattering if the $Z'$ is light but we have to use the full model. This automatically suppresses the signal of a sub-GeV $Z'$ in experiments with large momentum transfer $q^2 \gg \unit{GeV^2}$ such as NuTeV~\cite{Boehm:2004uq,Farzan:2015doa}, which otherwise provide strong bounds~\cite{Davidson:2003ha,Khan:2016uon}.
We see below that we can have large NSI without violating the constraints from Fig.~\ref{fig:neutrino_nucleon_scattering}.

\subsection{ Deviation from unitarity  and charged LFV processes \label{2.2}}

In the previous section we derived limits on $g_f/M_{Z'}$ for light $Z'$ coupled to $B$, $L$, or $B-L$.
In order to assess how large the NSI parameters from Eq.~\eqref{eq:ourNSI} can be, we further need to derive limits on the neutrino-mixing parameters $\kappa_\alpha$.
In the presence of the mixing between the SM neutrinos and $\Psi$, the Pontecorvo--Maki--Nakagawa--Sakata (PMNS) mixing matrix will deviate from unitarity. There are relatively strong bounds on the deviation from unitarity from various observables~\cite{Fernandez-Martinez:2016lgt}. Some of the bounds come from lepton flavor conserving observables such as muon decay or tests of lepton-flavor universality, which readily apply to our case, too:
\be
\label{Di} |\kappa_e|^2<2.5 \times 10^{-3},  \ |\kappa_\mu|^2<4.4 \times 10^{-4},  \ {\rm and} \  |\kappa_\tau|^2<5.6 \times 10^{-3} \ {\rm at } \ 2\sigma.
\ee
Note that not all limits from \emph{direct} searches for heavy neutrinos are applicable because our $\Psi$ decays mostly invisibly via $\Psi\to \nu Z'$ or $\Psi\to 3\nu$ for the parameters of interest. For $M_\Psi \gtrsim M_K$, this leaves us with direct-search bounds weaker than those from Eq.~\eqref{Di}, see Ref.~\cite{deGouvea:2015euy}. (We stress that our Dirac $\Psi$ does not contribute to $0\nu\beta\beta$.)
The bounds from LFV processes on the deviation from unitarity (i.e.~on $\kappa_\mu \kappa_\tau$,   $\kappa_\mu \kappa_e$ and  $\kappa_e \kappa_\tau$) found in Ref.~\cite{Fernandez-Martinez:2016lgt} do not apply to our case because in our model, the $\Psi$ state which mixes with $\nu$ can be much lighter than $M_W$ leading to the Glashow--Iliopoulos--Maiani (GIM) suppression of the corresponding contribution. Moreover, we have additional diagrams contributing to these rare LFV processes. We discuss the bounds from LFV in detail below. Before doing that, let us just notice that from Eq.~\eqref{Di} we obtain
\be \label{off-Di}  |\kappa_\mu \kappa_e|< 10^{-3}  \ , \ |\kappa_\mu \kappa_\tau|<1.6\times 10^{-3},  \ {\rm   and}\   |\kappa_e \kappa_\tau|<3.7\times 10^{-3}. \ee
It is remarkable that similar bounds still hold even if we increase the number of $\Psi$ because of the Cauchy--Schwarz inequality.
From this we can see that the NSI can still be large if $g_\Psi \gg g_f$ (for $f=u,d$, $g_f=g_B/3$ and $f=e$, $g_f=g_\ell$):
\begin{align}
\epsilon_{\alpha\beta}^{f} \simeq  \frac{g_f g_\Psi \kappa_\alpha^* \kappa_\beta}{2\sqrt2 G_F M_{Z'}^2}\simeq 0.3\times g_\Psi\left(\frac{\unit[1]{TeV}}{M_{Z'}/g_f}\right)\left(\frac{\unit[0.1]{GeV}}{M_{Z'}}\right) \left(\frac{\kappa_\alpha^* \kappa_\beta}{10^{-3}}\right) .
\end{align}
Depending on the gauge group, the matter NSI come from the coupling to electrons (for $U(1)_L$), neutrons and protons (for $U(1)_{B}$) or just neutrons (for $U(1)_{B-L}$, because the electron and proton $U(1)_{B-L}$ potentials cancel each other in neutral matter).

\begin{figure}[t]
\centering
\includegraphics[width=0.95\textwidth]{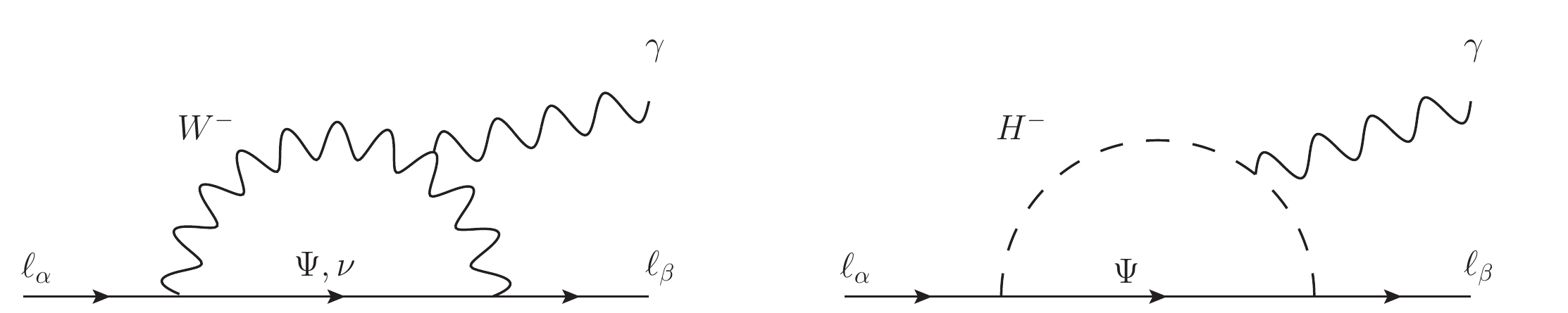}
\caption{
Loop-induced CLFV decay $\ell_\alpha \to \ell_\beta \gamma$.
}
\label{fig:radiativeLFV}
\end{figure}

Let us first discuss $l_\alpha \to l_\beta \gamma$. This process receives contributions from the $W^-$ loop and the $H^-$ loop (see Fig.~\ref{fig:radiativeLFV}). In the limit $m_\beta \ll m_\alpha \ll M_W$, $M_{H^-}$, we find the form factors
\begin{align}
F_{2R}^{W^-} &= \frac{g^2}{32\pi^2}\frac{m_\alpha^2}{M_W^2} \kappa_{\alpha }^* \kappa_{\beta } \left[ f_W (M_\Psi^2/M_W^2)- f_W(0)\right] ,\\
\label{F2H}
F_{2R}^{H^-} &= -\frac{1}{16\pi^2} \frac{m_\alpha^2}{M_{H^-}^2} y_\alpha^* y_\beta \sin^2\beta f_H (M_\Psi^2/M_{H^-}^2) \,,
\end{align}
which result in the  rate $\Gamma (\ell_\alpha\to \ell_\beta\gamma) = e^2 m_\alpha |F_{2R}^{W^-}+F_{2R}^{H^-}|^2/16\pi$, with loop functions~\cite{He:2002pva}
\begin{align}
f_W (x) &= \frac{10-43 x+78 x^2-49 x^3+4x^4 + 18 x^3 \log x}{12 (1-x)^4}\,,\\
f_H (x) &= \frac{1-6 x + 3 x^2 + 2 x^3 - 6 x^2 \log x}{12 (1-x)^4}\,.
\end{align}
Notice the GIM-like cancellation in $F_{2R}^{W^-}$ in the case of interest $M_\Psi\ll M_W$, which makes the $H^-$ contribution dominant for $\cos\beta \ll 1$,
\begin{align}
\Gamma (\ell_\alpha\to \ell_\beta\gamma) \simeq \frac{e^2}{4\pi}\,\frac{ |y_\alpha y_\beta|^2 m_\alpha^5}{384^2  \pi^4} \left(\frac{\sin^2\beta}{M_{H^-}^2} + \frac{3 \cos^2\beta}{M_W^2}\right)^2 .
\end{align}
From $\Br(\tau \to e \gamma)<3.3 \times 10^{-8}$ and $\Br(\tau \to \mu \gamma)<4.4 \times 10^{-8}$~\cite{Agashe:2014kda}, we then respectively obtain
\be \label{tauGamma}| y_e y_\tau |<0.46 \left(\frac{M_{H^-}}{\unit[400]{GeV}}\right)^2~~~~~{\rm and}~~~~~|y_\mu y_\tau| <0.53 \left(\frac{M_{H^-}}{\unit[400]{GeV}}\right)^2, \ee
 which can be readily satisfied. Similarly, from $\Br(\mu \to e \gamma)<4.2 \times 10^{-13}$~\cite{TheMEG:2016wtm} we obtain \be \label{eMu} |y_e y_\mu |< 7 \times 10^{-4} \left(\frac{M_{H^-}}{\unit[400]{GeV}}\right)^2.\ee
For values of $\kappa_\alpha$ satisfying Eq.~\eqref{off-Di}, the contribution from $F_{2R}^{W^-}$ to $\ell_\alpha\to \ell_\beta\gamma$
is negligible and well below the present bounds.

\begin{figure}[t]
\centering
\includegraphics[width=0.45\textwidth]{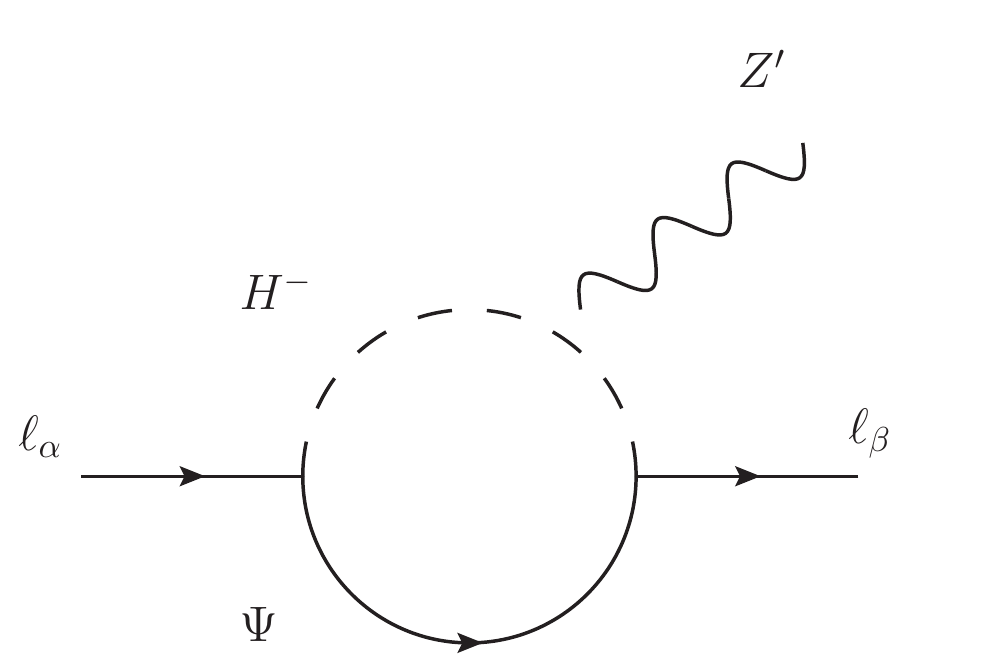}
\includegraphics[width=0.54\textwidth]{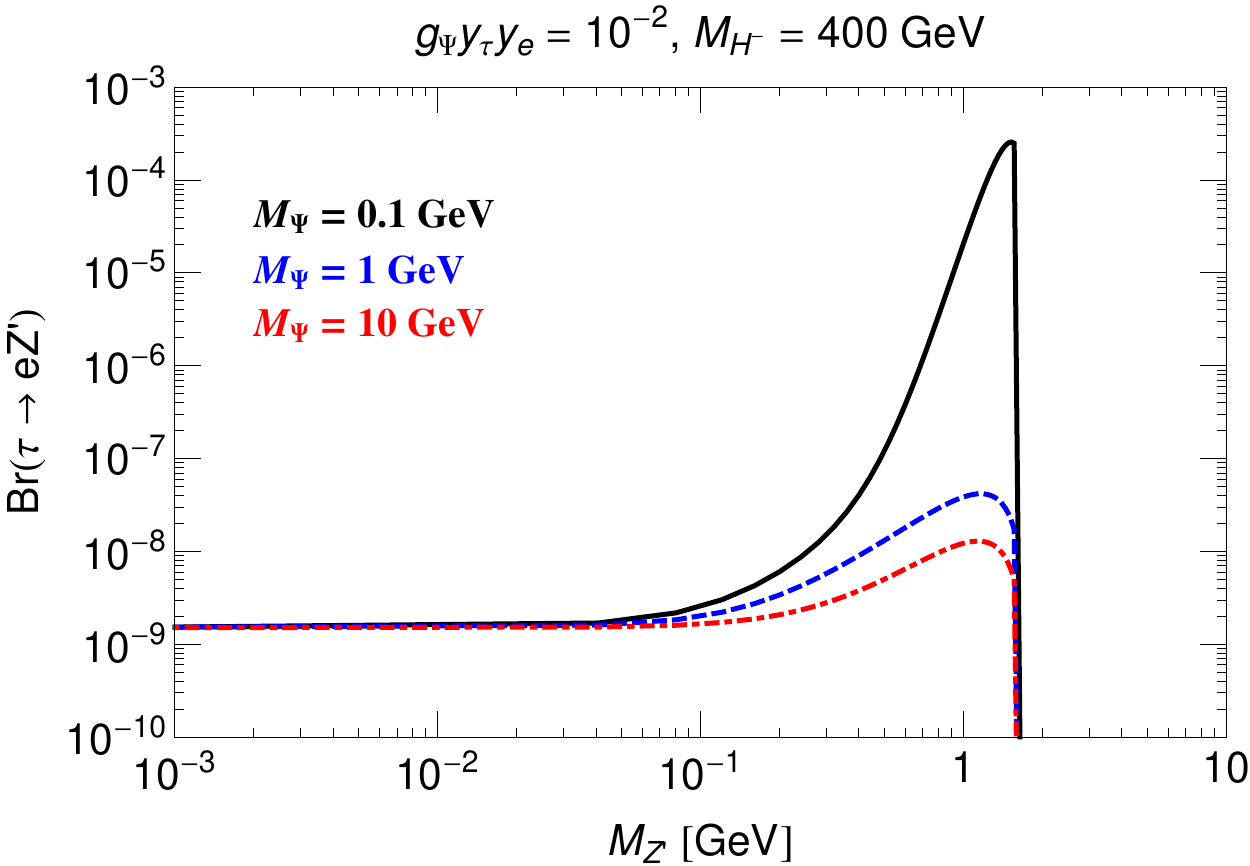}
\caption{
Left: Loop-induced CLFV decay $\ell_\alpha \to \ell_\beta Z'$. The $Z'$ can be attached  to both the $H^-$ and $\Psi$ lines.
Right: Resulting branching ratio $\Br (\tau\to e Z')$, which also holds for $\Br (\tau\to \mu Z')$ with $y_e\to y_\mu$, because we have  neglected the mass of  the  final fermion.
}
\label{fig:radiativeLFV_Zprime}
\end{figure}

On top of the $\ell_\alpha\to\ell_\beta\gamma$ constraints from above, there is CLFV involving the light~$Z'$. For $g_\ell = 0$, the $Z'$ only couples to charged leptons at one-loop level and so all processes $\ell_\alpha \to \ell_\beta \overline{\ell}_\gamma\ell_\delta$ will be two-loop suppressed.
Since we need a rather light $Z'$ to induce strong NSI, the main processes are the LFV two-body decays $\ell_\alpha \to \ell_\beta Z'$~\cite{Heeck:2016xkh}, followed by $Z'\to\nu\bar\nu$.
In the limit $m_\beta = 0$, the partial width $\ell_\alpha \to \ell_\beta Z'$ is given in terms of the two form factors of interest as
\begin{align}
\Gamma (\ell_\alpha \to \ell_\beta Z') = \frac{m_\alpha}{16 \pi }  \left[|F_{2,R}|^2 \left(1+\frac{M_{Z'}^2}{2 m_\alpha^2}\right)+|F_{1,L}|^2 \left(1  +\frac{m_\alpha^2}{2 M_{Z'}^2}\right)\right]\left(1-\frac{M_{Z'}^2}{m_\alpha^2}\right)^2 .
\end{align}
In the limit $\cos \beta \to 0$, the diagrams shown in Fig.~\ref{fig:radiativeLFV_Zprime} (left) give the dominant contribution,
\begin{align}
\begin{split}
F_{2,R} &= \frac{g_\Psi y_\alpha^* y_\beta}{64 \pi^2} \left(\frac{m_\alpha}{M_{H^-}}\right)^2  \left[\frac{x^2-1-2 x \log x }{ (x-1)^3}\right. \\
&\quad\left.+\frac{1}{3} \left(\frac{M_{Z'}}{M_{H^-}}\right)^2 \frac{  (1+x) \left(3-3 x^2+\left(1+4 x+x^2\right) \log x\right) }{  (x-1)^5} + \mathcal{O}\left(\frac{M_{Z'}^4}{M_{H^-}^4}\right)\right] ,
\end{split}
\end{align}
in which $x\equiv (M_\Psi/M_{H^-})^2$.
The charge-like form factor $F_1$ is also induced, albeit suppressed by the small gauge boson mass $M_{Z'}^2$,
\begin{align}
F_{1,L} =-\frac{g_\Psi y_\alpha^* y_\beta}{96 \pi^2}\left(\frac{M_{Z'}}{M_{H^-}}\right)^2 \left[ \frac{3-3 x+(2+x) \log x}{ (x-1)^2} \right]+ \mathcal{O}\left(\frac{M_{Z'}^4}{M_{H^-}^4}\right)\,.
\end{align}
As a result, the final decay rate is \emph{not} enhanced for $M_{Z'}\to 0$,\footnote{Compared to models where the $Z'_\mu$ couples at tree level to $\bar\ell_\alpha \gamma^\mu \ell_\beta$~\cite{Heeck:2014qea,Heeck:2016xkh}.} but rather goes smoothly to
\begin{align}
\Gamma (\ell_\alpha \to \ell_\beta Z') \xrightarrow{M_{Z'} \to 0} \frac{m_\alpha}{16 \pi } \left|\frac{g_\Psi y_\alpha y_\beta}{64 \pi^2}\right|^2 \left(\frac{m_\alpha}{M_{H^-}}\right)^4  \left[\frac{x^2-1-2 x \log x }{ (x-1)^3}\right]^2 ,
\end{align}
see Fig.~\ref{fig:radiativeLFV_Zprime} (right).
From $\Br(\tau \to e +{\rm light~ boson})<2.7\times 10^{-3}$ and  $\Br(\tau \to \mu +{\rm light~ boson})<5\times 10^{-3}$~\cite{Albrecht:1995ht}, we then find $|g_\Psi y_ey_\tau|<13\, (M_{H^-}/\unit[400]{GeV})^2$ and
 $|g_\Psi y_\mu y_\tau|<18 \,(M_{H^-}/\unit[400]{GeV})^2$ which can be readily satisfied and, using Eq.~\eqref{tauGamma}, provide a very weak bound on $g_\Psi$.
The contributions from the rest of the diagrams are suppressed by $\cos^2 \beta$ and as long as $g_\Psi y_\alpha y_\beta$ stays in the perturbative range cannot give rise to $\tau \to Z^\prime \mu (e)$ rates above the bounds.
Similarly, $\Br(\mu \to e Z^\prime)\simeq 8.6 \times 10^{-5} |g_\Psi y_\mu y_e|^2 (\unit[400]{GeV}/M_{H^-})^4$ for light $Z'$, which for $y_ey_\mu$ satisfying Eq.~\eqref{eMu} is much lower than even the strongest bounds from rare muon decay modes. (Limits on $\Br(\mu \to e +{\rm light~ boson})$ are of order $10^{-5}$~\cite{Jodidio:1986mz,Bayes:2014lxz}.)

The above discussion shows that the $\ell_\alpha \to \ell_\beta Z'$ constraints are weaker than those from $\ell_\alpha \to \ell_\beta \gamma$, even for large $g_\Psi$. For $g_\ell \neq 0$, however, the $Z'$ mediated $\ell_\alpha \to \ell_\beta \overline{\ell}_\delta\ell_\delta$ could be enhanced. (For $g_B \neq 0$, the decays are $\ell_\alpha \to \ell_\beta \pi^+\pi^-$, $\ell_\beta \pi^0 \gamma$, $\ell_\beta \pi^+\pi^-\pi^0$, which are much less constrained.)
Focusing on the region of parameters with $2m_e <M_{Z'} < m_\tau-m_\mu, M_\Psi$, $\kappa_\tau \kappa_\beta\neq 0$, the resonantly enhanced LFV decay rate $\tau\to \ell_\beta \overline{\ell}_\delta\ell_\delta$ can be estimated as
\begin{align}
\Gamma (\tau\to \ell_\beta \overline{\ell}_\delta\ell_\delta) \simeq \Gamma (\tau\to \ell_\beta Z') \Br (Z'\to\overline{\ell}_\delta\ell_\delta)\,,
\end{align}
with
\begin{align}
\Br (Z'\to\overline{\ell}_\delta\ell_\delta) \simeq \frac{g_\ell^2}{2g_\ell^2+  \tfrac12 \sum_{\alpha,\beta} |g_\ell \delta_{\alpha\beta} + g_\Psi\kappa_\alpha^*\kappa_\beta|^2} \,,
\end{align}
neglecting fermion masses. Even if this branching ratio is of order one, the total $\Gamma (\tau\to \ell_\beta \overline{\ell}_\delta\ell_\delta)$ is still suppressed by the loop factor and the potentially large $M_{H^-}$, so it is not necessarily dangerous.
From Fig.~\ref{fig:radiativeLFV_Zprime} (right) we see that values $|g_\Psi y_\alpha y_\beta |<10^{-2}$ can suppress these decays below the experimental limits of $\mathcal{O}(10^{-8})$~\cite{Agashe:2014kda} (conservatively assuming that the $Z'$ decay is prompt and does not lead to a secondary vertex). We show this most pessimistic constraint of $\Br (\tau\to \ell_\beta Z') \lesssim \mathcal{O}(10^{-8})$ in Fig.~\ref{fig:NSIcontoursBL}, but stress again that it is expected to be much weaker.
Similar conclusions hold for the other off-diagonal NSI. For \emph{diagonal} NSI, no limits from CLFV arise and the strongest limits come from neutrino scattering.

\begin{figure}[t]
\centering
\includegraphics[width=0.49\textwidth]{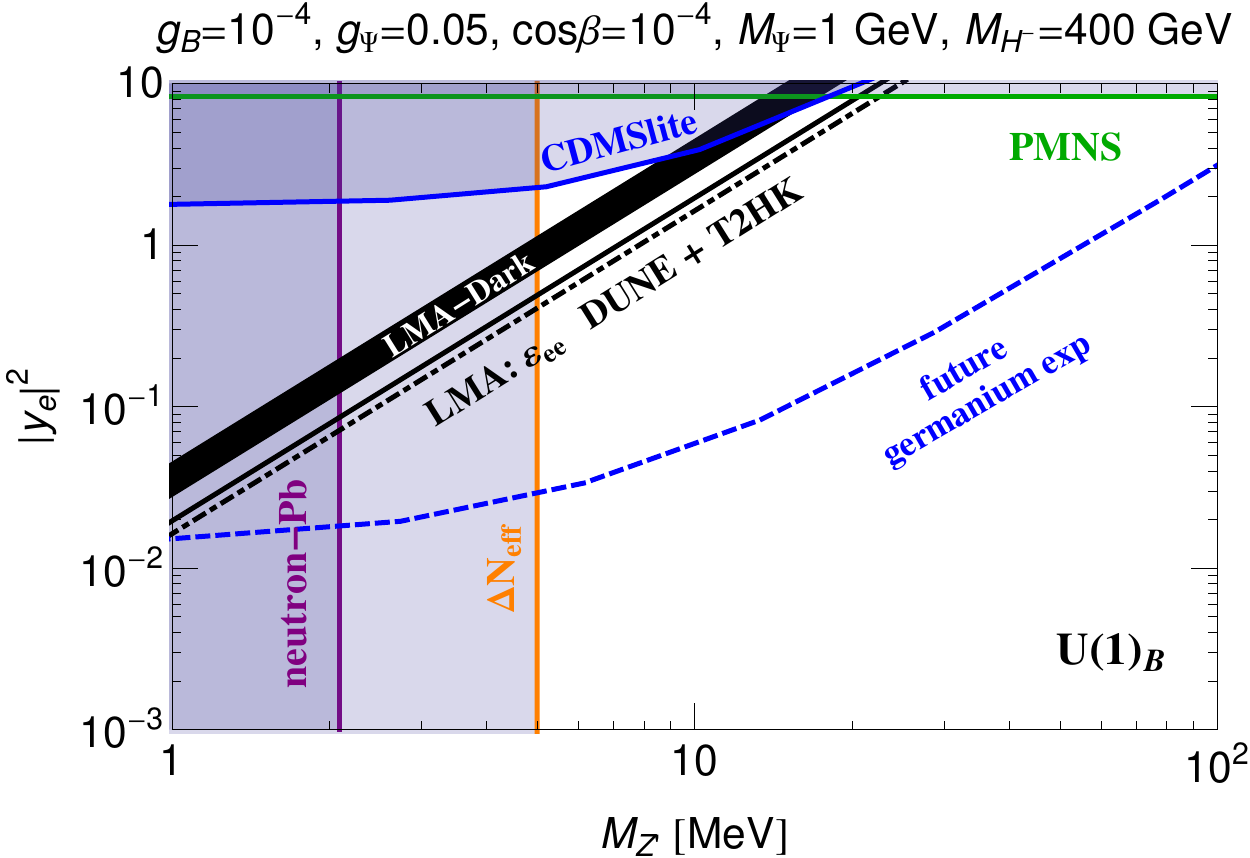}
\includegraphics[width=0.49\textwidth]{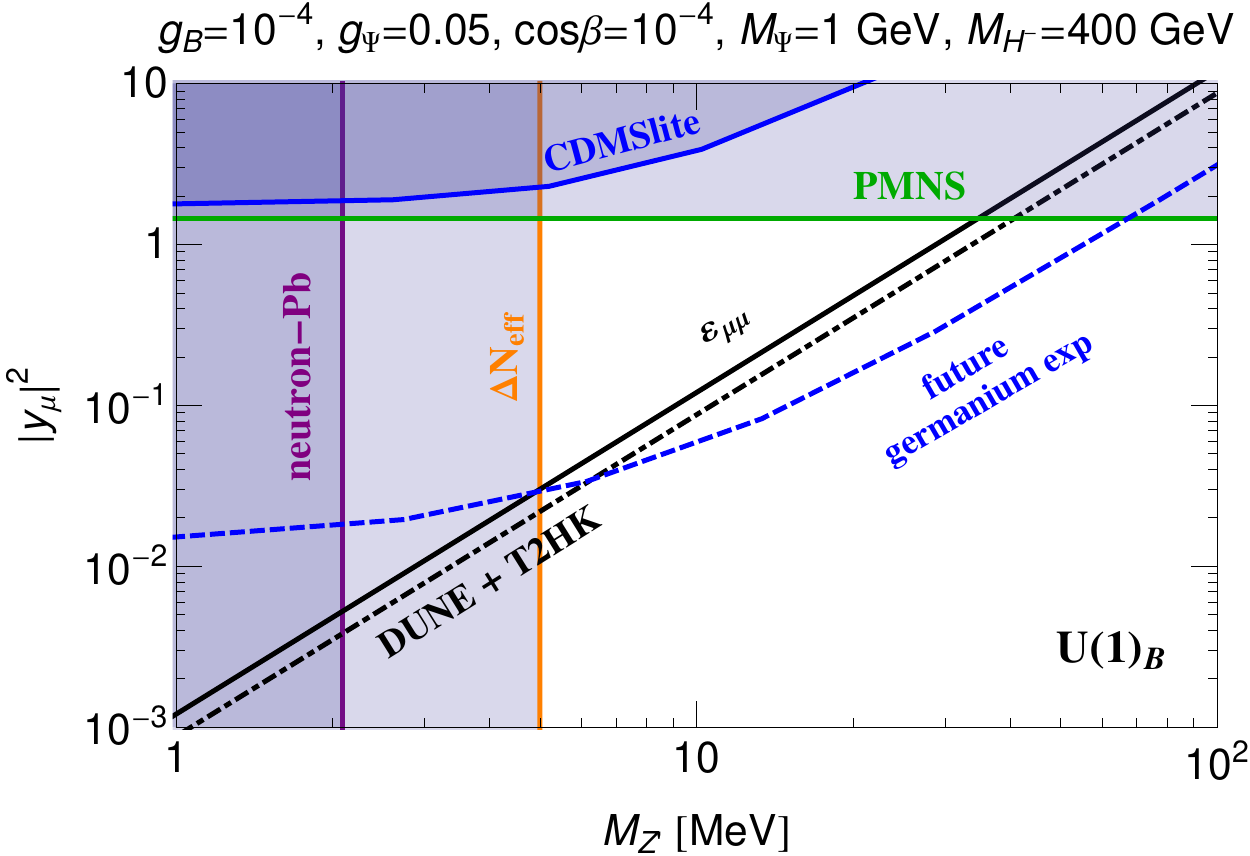}
\includegraphics[width=0.49\textwidth]{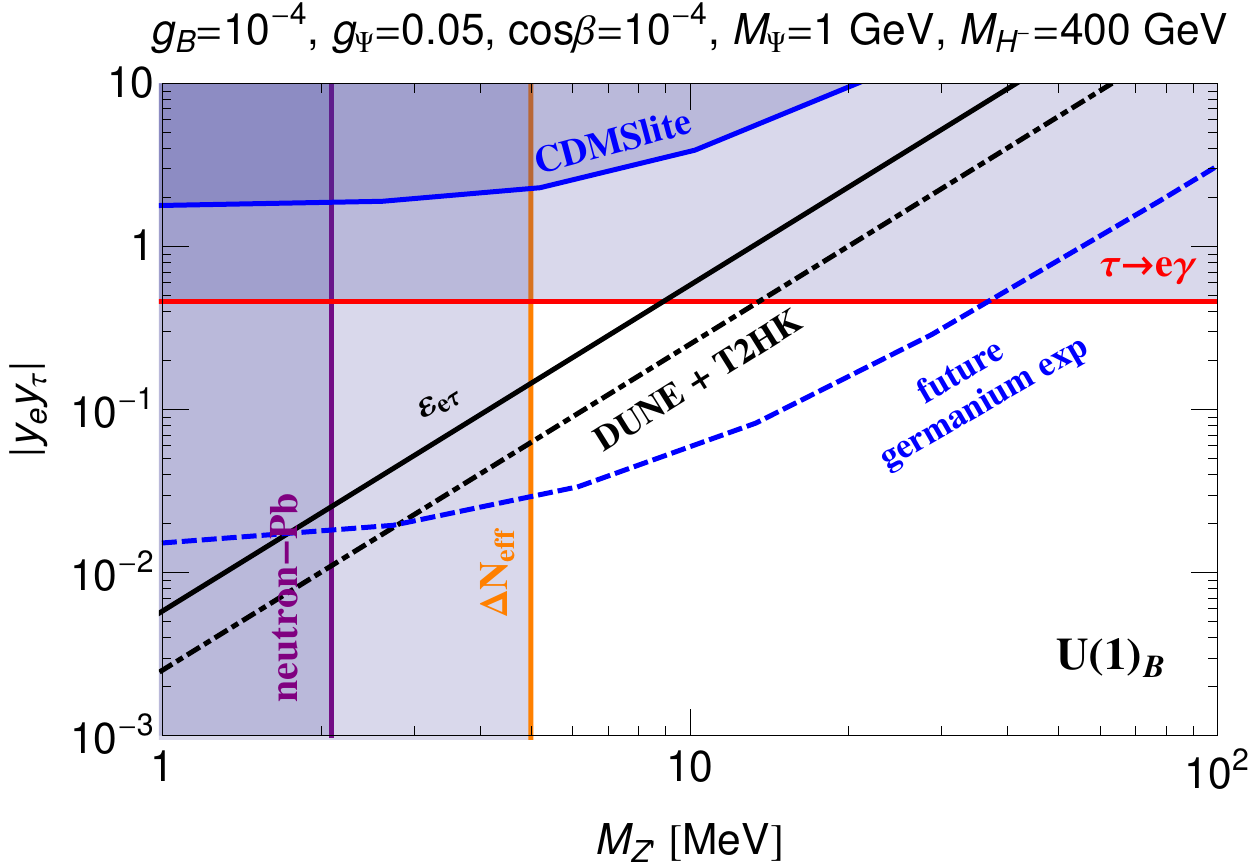}
\includegraphics[width=0.49\textwidth]{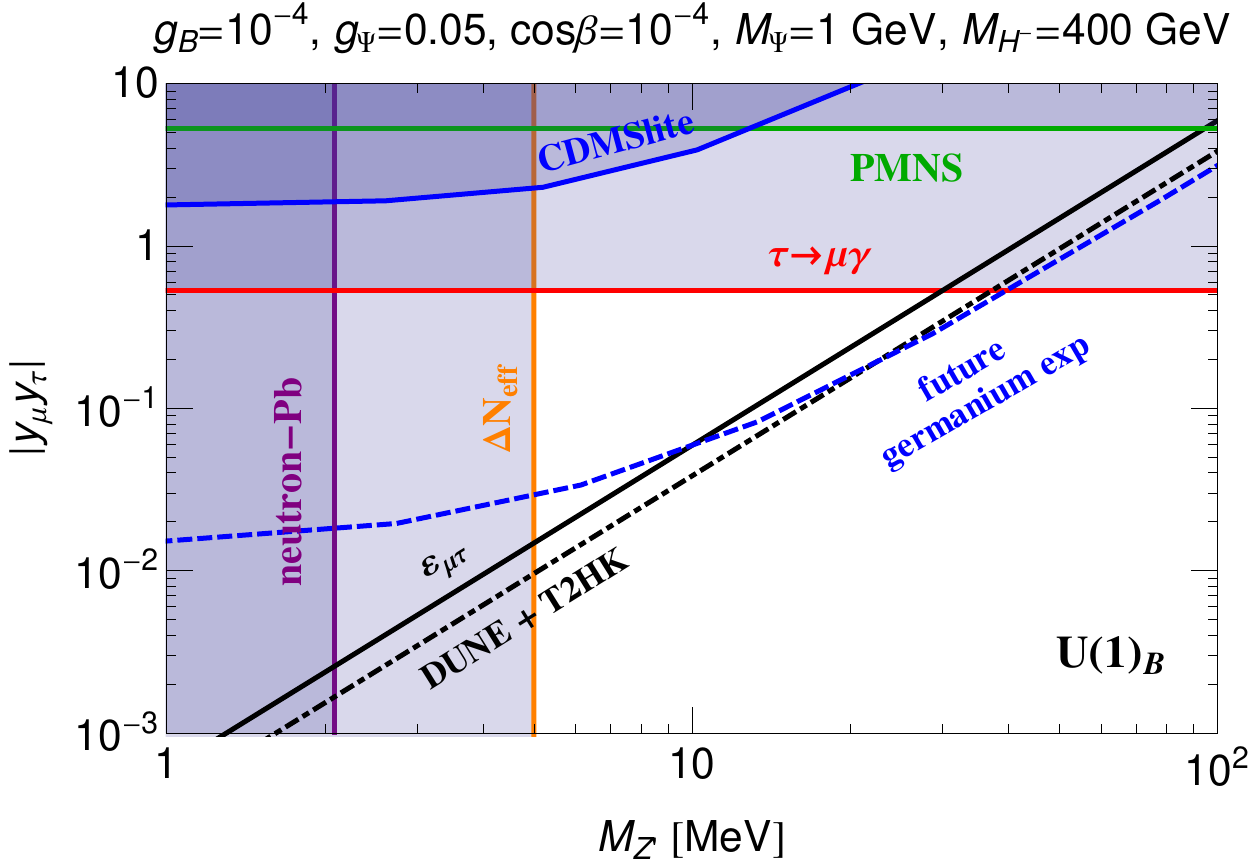}
\caption{
 Contours of diagonal NSI (top left and right) and off-diagonal NSI (bottom left and right)  for $U(1)^\prime=U(1)_B$. Relevant experimental bounds are also shown (see the text for details). The inequality of Eq.~\eqref{BoundsOnbeta} is satisfied in the allowed parameter space.
The current (solid black lines) and projected (dot-dashed black line marked with DUNE+T2HK, obtained by including a prior for current constraints) limits on $\epsilon \simeq \epsilon^e+3\epsilon^u +3\epsilon^d$ are taken from Ref.~\cite{Coloma:2015kiu}. For $\epsilon_{ee}$ (top left) we also indicate the preferred region for the LMA-Dark solution.
}
\label{fig:NSIcontoursB}
\end{figure}

Putting everything together, we can illustrate the size of our NSI for some benchmark points in connection to the other constraints. In Fig.~\ref{fig:NSIcontoursB} we show the least-constrained case: $U(1)_B$. The resulting NSI from Eq.~\eqref{eq:ourNSI} can be of order one without being in conflict with any of the other constraints, even for the off-diagonal LFV NSI.\footnote{
An exception is $\epsilon_{e\mu}$, which is typically tiny to satisfy $\mu\to e\gamma$, unless $H^-$ is very heavy (Eq.~\eqref{eMu}).}
 The best constraints then come from the actual neutrino-oscillation experiments, and are improved e.g.~with DUNE. In particular, the LMA-Dark solution can be realized (see Sec.~\ref{implications}). Notice that we have drawn contour plots for a combination of $\epsilon^f$ that is relevant for propagation in Earth with fermion densities $n_n/n_e \simeq n_p/n_e=1$ (see Sec.~\ref{implications}).
The vertical line at \unit[5]{MeV} is the lower bound from cosmology on $M_{Z^\prime}$ under conservative assumption $\Delta N_\text{eff}<0.7$~\cite{Ahlgren:2013wba,Kamada:2015era}. Current (approximate) limits on solar-neutrino--nucleus scattering from CDMSlite are only relevant for large $\epsilon\sim 1$, i.e.~only for the LMA-Dark solution. Future germanium or xenon experiments for dark matter detection such as SuperCDMS SNOLAB and LUX--ZEPLIN will however provide a powerful method to test our model via nuclear recoils~\cite{Cerdeno:2016sfi}, competitive with DUNE and T2HK. Here we have again ignored the flavor composition of solar neutrinos.

\begin{figure}[t]
\centering
\includegraphics[width=0.49\textwidth]{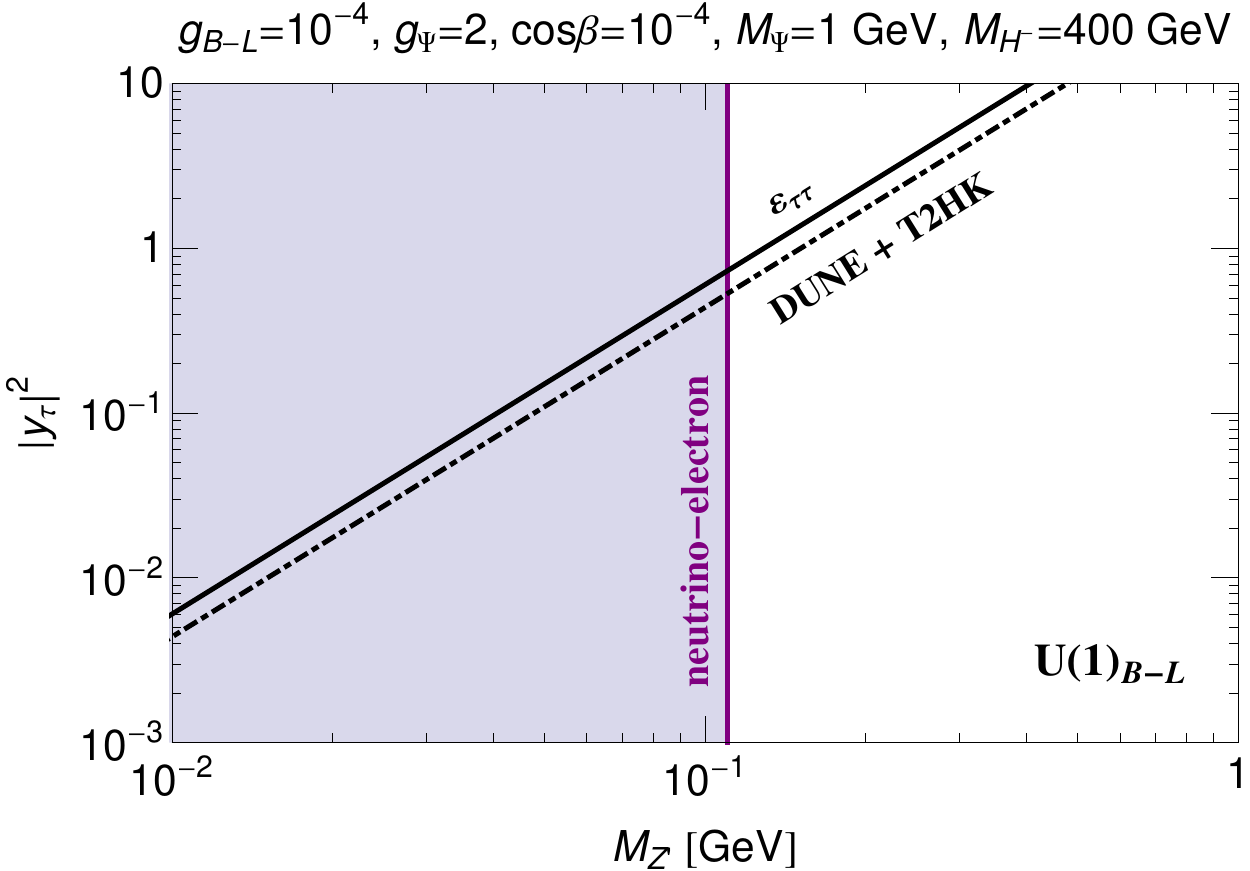}
\includegraphics[width=0.49\textwidth]{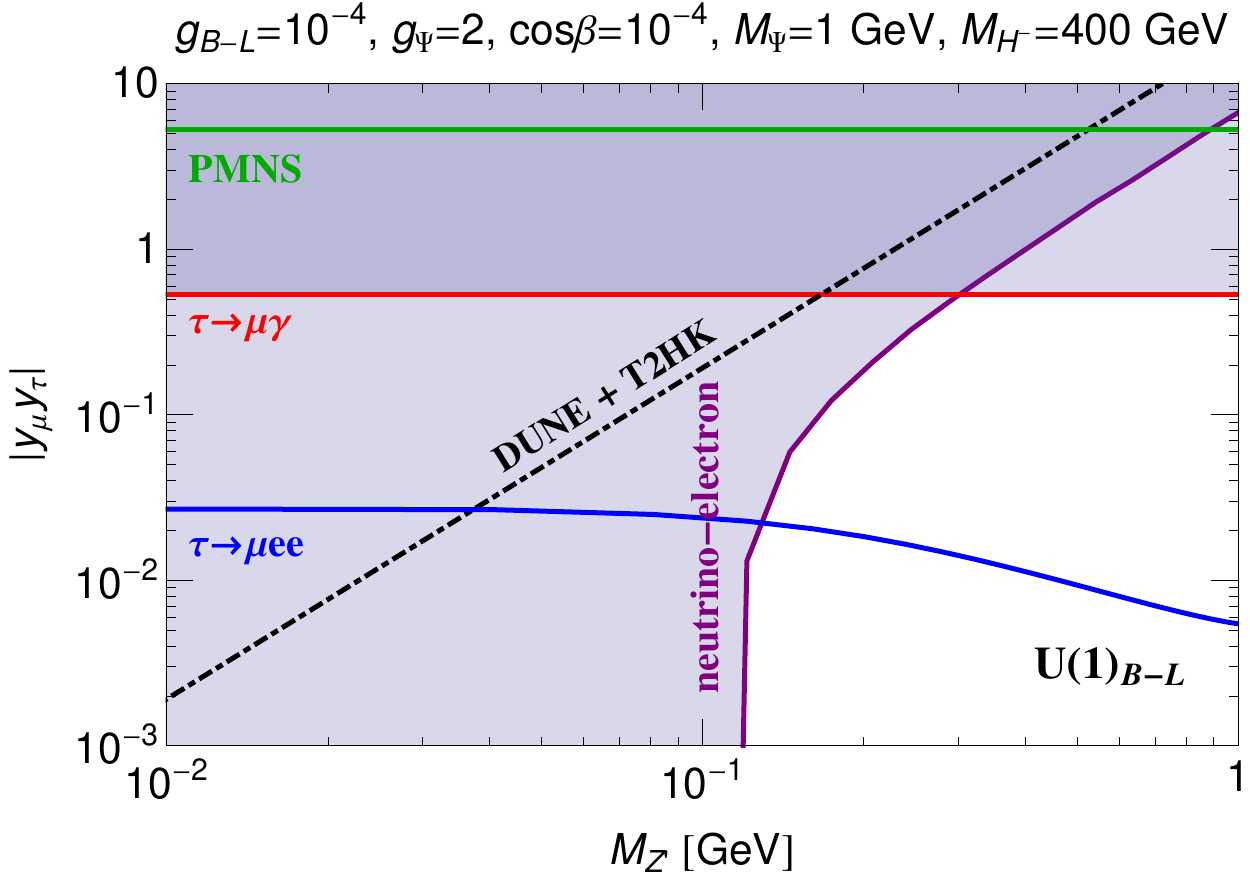}
\caption{
Contours of diagonal NSI ($\epsilon_{\tau\tau}$, left) and off-diagonal NSI ($\epsilon_{\mu\tau}$, right) and additional constraints for a set of parameters of $U(1)_{B-L}$. The inequality of Eq.~\eqref{BoundsOnbeta} is satisfied in the allowed parameter space.
The constraint from $\tau\to\mu ee$ is extremely conservative.
}
\label{fig:NSIcontoursBL}
\end{figure}

Taking instead $U(1)_{B-L}$ as our gauge group gives a much more restricted picture (Fig.~\ref{fig:NSIcontoursBL}). It is not possible to generate $\epsilon_{ee}\sim 1$ due to the strong constraints from neutrino--electron scattering experiments, so the LMA-Dark solution is incompatible with $B-L$. Large diagonal NSI can only be obtained for the $\tau\tau$ entry (Fig.~\ref{fig:NSIcontoursBL} (left)), because the additional $Z'\nu_\tau\nu_\tau$ coupling is the only one not constrained by neutrino--electron scattering experiments (which only use $\nu_e$ and $\nu_\mu$).
{Note that $\epsilon_{\tau\tau}$ contributes to two parameters in the analysis of Ref.~\cite{Coloma:2015kiu}, $\tilde \epsilon_{ee}\equiv \epsilon_{ee}-\epsilon_{\tau\tau}$ and $\tilde \epsilon_{\mu\mu}\equiv \epsilon_{\mu\mu}-\epsilon_{\tau\tau}$; in Fig.~\ref{fig:NSIcontoursBL} we show the bound and DUNE projection from $\tilde \epsilon_{\mu\mu}$, which is the stronger of the two.
}
For the off-diagonal LFV NSI, the additional limits from $\ell_\alpha\to \ell_\beta \overline{\ell}_\delta\ell_\delta$ severely restrict the parameter space compared to $U(1)_B$ (Fig.~\ref{fig:NSIcontoursBL} (right)). We could evade those bounds only by increasing  $M_{H^-}$, we would then still face the bounds from neutrino scattering, as all off-diagonal NSI involve either $\nu_e$ or $\nu_\mu$. In view of this, the $B-L$ model is a simple framework to generate a sizable $\epsilon^{u,d,e}_{\tau\tau}$, but all other NSI are typically restricted to be tiny.
{We stress again that it is in principle possible to severely suppress the $Z'$ couplings to protons and charged leptons by tuning the kinetic-mixing angle to cancel the $B-L$ coupling~\cite{Heeck:2014zfa}, thus weakening the constraints without affecting the NSI, which come from the coupling to \emph{neutrons}.}

\subsection{UV completion \label{UV}}

 Let us outline the UV completion of our models. The components of the new scalar doublet $H^\prime$ (having electroweak interactions) should be heavier than the electroweak scale. On the other hand,
 $\langle H^\prime \rangle =v\cos \beta/\sqrt{2} \ll v $ (see Eq.~\eqref{BoundsOnbeta}). This can be obtained by introducing a new singlet scalar $S_1$ with $U(1)^\prime$ charge equal to that of $H'$. We can then add the following trilinear term to the Lagrangian
\begin{align}
\L = \mu S_1^\dagger H^\dagger H' \,,
\end{align}
which results in a softly broken 2HDM after $S_1$ acquires a VEV~\cite{Heeck:2014qea}. The VEV $\langle H'\rangle\simeq -\mu \langle S_1 \rangle \langle H\rangle /(2M_{H^\prime}^2)$  is induced by $\langle S_1\rangle$ without creating any massless Goldstone bosons. Notice that $\langle S_1 \rangle \sim M_{S_1}$ can be much lighter than the electroweak scale. Taking $\langle S_1 \rangle \mu\ll M_{H'}^2$, we can naturally obtain $\cos \beta\ll 1$.
Note that the details of the scalar potential are not important for the NSI phenomenology.
	
Let us discuss the implications for our different gauge symmetries:
\begin{itemize}
\item
$U(1)^\prime=U(1)_{B-L}$ or $U(1)_L$:
  In order to employ the seesaw mechanism and make the right-handed neutrinos sufficiently heavy -- otherwise Big Bang nucleosynthesis would kill our light-$Z'$ parameter space~\cite{Heeck:2014zfa} -- we need a scalar $S_R\sim -2g_\ell$ to couple $ S_R \bar\nu_R^c \nu_R \to \mathcal{M}_R \bar\nu_R^c \nu_R$. Unfortunately, the scalar potential $V(H,H',S_1,S_R)$ has an additional global $U(1)$ symmetry that results in a Goldstone boson when all fields acquire VEVs. This can be avoided by introducing yet another scalar $S_2~\sim g_\Psi + g_\ell$ that couples $S_1 S_R^\dagger S_2^\dagger$ and breaks the unwelcome global symmetry explicitly.

\item
$U(1)^\prime=U(1)_{B}$:
	 Here the $U(1)^\prime$ charge of the neutrinos is zero, so a Majorana mass term for $\nu_R$ is allowed by symmetry and the seesaw mechanism works without problems. However, we have to introduce new particles $\eta$ that cancel the $U(1)_B\times [SU(2)_L\times U(1)_Y]$ gauge anomalies from the quarks.
The simplest realizations introduce two ``lepton'' generations with charges $B_1$ and $B_2$, which cancel the anomalies for $B_1-B_2 = -3$~\cite{Duerr:2013dza,Duerr:2014wra}, or two doublets, one triplet and one singlet~\cite{Perez:2014qfa,Ohmer:2015lxa}.
 These new particles are non-chiral under the electroweak gauge group but chiral under the $U(1)^\prime$. Hence, they can obtain a mass above the electroweak scale by the VEV of an electroweak singlet scalar $S_B$ with $\langle S_B\rangle \sim M_\eta\gg \unit[100]{GeV}$, where $M_\eta$ is the typical mass scale of the new $\eta$ particle. (To avoid Goldstone bosons, a third scalar $S_2$ is typically required in our model as well.) This VEV $\langle S_B\rangle$ induces a mass for $M_{Z'}$ given by $g_B \langle S_B\rangle$. Taking $\langle S_B\rangle\gtrsim \unit[1]{TeV}$, we find
\begin{align}
 \label{gBbound}
 g_B \lesssim 10^{-4} \left(\frac{M_{Z'}}{\unit[100]{MeV}}\right)
\end{align}
as an additional rough bound required to make the anomaly-canceling fermions sufficiently heavy. Note that this does however depend strongly on the detailed mass spectrum and mixing pattern of the $\eta$ fermions.
To make matters even more involved, the lightest of the $\eta$ particles is stabilized by the remaining unbroken $\mathbb{Z}_3^B$ subgroup, and thus forms dark matter~\cite{Duerr:2014wra,Ohmer:2015lxa}. This further complicates the question of how large $M_\eta$ and thus $\langle S_B\rangle$ have to be to avoid constraints from collider searches and direct-detection experiments. A discussion of the dark matter sector goes unfortunately beyond the scope of this article and is left for future work.

\item $U(1)' = U(1)_{B-\sum_\alpha x_\alpha L_\alpha}$: Having focused on flavor-universal couplings to $B$ and $L = L_e+L_\mu+L_\tau$ so far, let us briefly mention the possibility of (lepton-)flavored gauge symmetries. The symmetry $U(1)_{B-\sum_\alpha x_\alpha L_\alpha}$ is anomaly free for $\sum_\alpha x_\alpha = 3$, which reduces to $B-L$ for $x_{e,\mu,\tau} = 1$. For $x_\alpha\neq x_\beta$, the $Z'$ couplings break lepton universality and lead to different phenomenology compared to $B-L$. The case of interest here is $x_e = 0$, in order to eliminate the strong constraints from electron-scattering experiments. For $x_\mu\neq 0$, this still leaves (significantly weaker) constraints from the anomalous magnetic moment of the muon or $\nu_\mu$ scattering~\cite{Altmannshofer:2014pba}. The extreme case $B-3L_\tau$~\cite{Ma:1997nq,Ma:1998dp} is almost as weakly constrained as $U(1)_B$ and therefore perfectly suited to generate large NSI. (Even without the introduction of $\Psi$ it would lead to non-zero $\epsilon^{u,d}_{\tau\tau}$.)
For these flavored $U(1)'$ we do not have to worry about anomaly-canceling fermions -- as in the $U(1)_B$ case -- but instead about how to obtain the observed leptonic mixing pattern, i.e.~the PMNS matrix. As shown in Ref.~\cite{Araki:2012ip}, it requires only a few singlet scalars to generate viable neutrino mass matrices for these $U(1)'$ via seesaw, and could easily lead to predictions in the form of texture zeros or vanishing minors.

\end{itemize}

\section{Phenomenological implications of our model}
\label{implications}

In the previous section we constructed a viable model that can give rise to large neutral-current NSI for neutrinos.
In this section, we discuss the phenomenological implications of the model for oscillation experiments, direct searches of $H^-$ at colliders and charged lepton dipole moments.

Neutrino propagation in matter is sensitive to a combination of couplings to different fermions weighted by their density $n_f$ in that particular medium,
\begin{align}
\epsilon^m_{\alpha\beta} \equiv \sum_{f} \frac{n_f}{n_e}\,\epsilon_{\alpha\beta}^{f}\,.
\end{align}
In electrically neutral matter, the electron and proton densities are equal, $n_e = n_p$, so we can simplify this to
\begin{align}
\epsilon^m_{\alpha\beta} = \left(\epsilon_{\alpha\beta}^{e} + 2 \epsilon_{\alpha\beta}^u + \epsilon_{\alpha\beta}^d\right)+ \frac{n_n}{n_e}\,\left(  \epsilon_{\alpha\beta}^u + 2\epsilon_{\alpha\beta}^d\right) ,
\label{eq:neutralmatterNSI}
\end{align}
with the neutron density $n_n$.
As already stated above, a $Z'$ coupling to baryon number gives $\epsilon^e = 0\neq\epsilon^u=\epsilon^d$, whereas a coupling to lepton number gives $\epsilon^e \neq 0=\epsilon^u=\epsilon^d$. An interesting special case arises for $B-L$, as in that case $\epsilon^u/3=\epsilon^d/3 = -\epsilon^e$, so the first bracket in Eq.~\eqref{eq:neutralmatterNSI} vanishes, leaving only a coupling to \emph{neutrons}. This has an impact on neutrino oscillations inside the Sun, not only because the total number of neutrons is smaller than that of protons and electrons, but the neutron density has a different spatial dependence (tracing essentially the Helium abundance, peaked towards the core~\cite{Bahcall:2000nu}).
We are not aware of an NSI analysis under this condition.

Let us discuss the possibility of reproducing the famous LMA-Dark solution with $\theta_{12}>45^\circ$ and $\epsilon_{\mu \mu}^q-\epsilon_{ee}^q\simeq \epsilon_{\tau \tau}^q-\epsilon_{ee}^q\simeq 1$,
which seems to provide an even better fit to the solar neutrino data than the standard LMA solution \cite{Gonzalez-Garcia:2013usa,Miranda:2004nb,Escrihuela:2009up}. It was shown in Ref.~\cite{Gonzalez-Garcia:2013usa} that this solution survives global oscillation tests, provided that off-diagonal elements of $\epsilon$ as well as the splitting between $\epsilon_{\mu\mu}$ and $\epsilon_{\tau \tau}$ satisfy relatively stringent upper bounds of order of 0.01--0.1. Ref.~\cite{Farzan:2015doa} presented a model that gave rise to the LMA-Dark solution with $\epsilon_{\mu \mu}^q=\epsilon_{\tau\tau}^q\simeq 1$ and $\epsilon_{ee}^q= \epsilon_{\alpha \beta} |_{\alpha \ne \beta}^q=0$. The present model which gauges $U(1)_B$ (but not $U(1)_{B-L}$) can also provide a theoretical foundation for the LMA-Dark solution: taking
$\kappa_\mu=\kappa_\tau=0$,
\begin{align}
 |\kappa_e|^2 \sim 10^{-3} \left| \frac{1}{g_\Psi} \times \frac{10^{-5}}{g_B}\right| \left(\frac{M_{Z'}}{\unit[10]{MeV}}\right)^2 ,
\end{align}
and $g_\Psi g_B<0$, we can reproduce the range of values compatible with the LMA-Dark solution. Notice that unlike in the model of Ref.~\cite{Farzan:2015doa}, here we have
$\epsilon_{ee}^{u,d}\sim -1$ and $\epsilon_{\mu\mu}^{u,d}=\epsilon_{\tau\tau}^{u,d}=0$, which is equivalent to $\epsilon_{ee}^{u,d} = 0$ and $\epsilon_{\mu\mu}^{u,d}=\epsilon_{\tau\tau}^{u,d}\sim 1$ in neutrino propagation.
(Generating directly $\epsilon_{\mu \mu}^{u,d}=\epsilon_{\tau \tau}^{u,d}\sim 1$ in our model implies $|\epsilon_{\mu \tau}^{u,d}|\sim 1$, which is not compatible with atmospheric neutrino data~\cite{Gonzalez-Garcia:2013usa}.)

Apart from the LMA-Dark solution, no other solution with preferred nonzero $\epsilon$ has been found. Within the standard neutrino-oscillation paradigm with $\theta_{12}<45^\circ$, relatively strong upper bounds are set on the values of $|\epsilon_{\alpha \beta}|_{\alpha \ne \beta}$ and $|\epsilon_{\alpha \alpha}-\epsilon_{\beta \beta}|$.  Within our $U(1)_B$ model as shown in Fig.~\ref{fig:NSIcontoursB}, these bounds can be easily saturated without being in conflict with any other observational bound. For a generic flavor pattern of $\epsilon_{\alpha \beta}$, if the values of $\epsilon_{\alpha \beta}$ are close to these bounds, the upcoming long-baseline experiments will be able to probe $\epsilon$. In specific cases when certain relations hold among the values of $\epsilon_{\alpha \beta}$, the effects of NSI hide from observation (see {\it e.g.,} Ref.~\cite{Yasuda:2015lwa} and Fig.~4 of Ref.~\cite{Coloma:2015kiu}). As discussed before, the relation that our model predicts is $|\epsilon_{\alpha \beta}|^2=\epsilon_{\alpha \alpha}\epsilon_{\beta \beta}$. For such relations, NSI effects at long-baseline experiments can be observable.

As shown in the literature, the phases of $\epsilon_{\alpha \beta}|_{\alpha \ne \beta}$ can have important effects on the DUNE experiment and can introduce new degeneracies~\cite{Agarwalla:2016fkh}.  The phase of $\epsilon_{\alpha \beta}$ can originate from the mismatch of the phases of $y_\alpha$ and $y_\beta$. If the phases of all $y_\alpha$ are the same, $y_\alpha$ can of course be made real by rephasing $\Psi$. However, for $\Im[y_\alpha /y_\beta]\ne 0$,  this is not possible. The phase of $y_\alpha$ can be absorbed by rephasing $L_\alpha$ but the phase reappears in the neutrino mass matrix. For experiments such as long-baseline neutrino experiments where the tiny neutrino masses have observable effects, this phase can also have an observable effect, but for the electric dipole moment of charged leptons,  the effects of this new source of CP violation are suppressed by the neutrino mass and are therefore negligible.

Lepton-flavor conserving diagrams similar to those of Fig.~\ref{fig:radiativeLFV} contribute to lepton magnetic moments.
Setting $\alpha =\beta$ in Eq.~\eqref{F2H}, we can calculate the contribution of $H^-$ coupling to the magnetic dipole moment of the charged leptons as
\be
\delta a_\mu\sim 10^{-12} \left| \frac{y_\mu}{0.1}\right|^2 \left( \frac{\unit[300]{GeV}}{M_{H^-}}\right)^2 ~~~{\rm and}~~~
\delta a_e\sim 10^{-16} \left| \frac{y_e}{0.1}\right|^2 \left(\frac{\unit[300]{GeV}}{M_{H^-}}\right)^2 , \ee
which are both below the uncertainty in the measurements of these quantities~\cite{Agashe:2014kda}.

One of the essential ingredients of the present model is the presence of a charged scalar $H^-$ which dominantly decays to a charged lepton plus $\Psi$ with branching ratios
\begin{align}
\frac{\Br (H^-\to \ell_\alpha \Psi)}{\Br (H^-\to \ell_\beta \Psi)}\simeq \frac{|y_\alpha|^2}{|y_\beta|^2} \simeq \frac{\epsilon_{\alpha\alpha}}{\epsilon_{\beta\beta}} \,.
\end{align}
Since the main decay decay mode of $\Psi$ is $\Psi \to \nu Z^\prime$ and $Z^\prime$ subsequently decays into a neutrino pair, $\Psi$ should appear as missing energy at colliders.  If the mass of $H^-$ is smaller than the beam energy at the LHC, the $H^-H^+$ pairs  can be produced by electroweak interactions ({\it i.e.,} $s$-channel $\gamma/Z$ exchange) and their decay leads to dilepton plus missing energy signal. The discovery potential of the LHC is explored in  Ref.~\cite{Farzan:2010fw}.  The strongest bound on such a charged scalar is still provided by the LEP experiments, $M_{H^-}>\unit[90]{GeV}$, assuming $\Br(H^-\to \tau\nu) = 1$~\cite{Abbiendi:2013hk}.
{We are not aware of similar searches for $H^-\to e\nu$ or $\mu\nu$ by LEP, which could easily be dominant in our case.} However the signature of $H^-$ at the LHC with $y_e \ne 0$ and $y_\mu=y_\tau=0$ (with $y_\mu \ne 0$ and $y_e =y_\tau=0$) is very similar to the signature of a left-handed selectron (left-handed smuon) decaying to the electron (muon) plus a light neutralino. The CMS collaboration using $L=\unit[19.5]{fb^{-1}}$ and $\sqrt{s}=\unit[8]{TeV}$ data has ruled out such a charged scalar with mass in the range 125--\unit[275]{GeV} (see Fig.~14 of \cite{Khachatryan:2014qwa}). If $H^-$ decaying to $\mu$ and $e$ was lighter than \unit[125]{GeV}, it should have been discovered at LEP as the reconstruction of the muon or the electron is much simpler than the $\tau$ lepton. Thus, it is safe to claim $M_{H^-}>\unit[275]{GeV}$ for $y_e\ne 0$ or $y_\mu\ne 0$.  If $y_e$ is comparable to electroweak couplings
({\it e.g.,} to reproduce the LMA-Dark solution), along with $s$-channel $Z/\gamma$ exchange, the $t$-channel $H^-H^+$ pair production at ILC via $\Psi$ exchange  can also be important  provided that $s>4 M_{H^-}^2$. The signal is an excess in $e^-e^+$+ missing energy relative to the SM prediction.

In the model which gauges $U(1)_B$, the coupling of the $Z^\prime$ boson to $\bar{\nu}_\beta \gamma^\mu \nu_\alpha$ current can be estimated as \be\label{nuco}
|g_\Psi \kappa_\alpha \kappa_\beta|\sim  10^{-4}\, \epsilon_{\alpha \beta}^{u,d} \left(\frac{M_{Z^\prime}}{\unit[10]{MeV}}\right)^2 \left( \frac{10^{-4}}{g_B}\right).
\ee
The new interaction can give rise to correction to meson decay $K^-,\pi^-\to \ell_\alpha^- + Z^\prime +\nu_\beta$, which appears as a new contribution to $K$ or $\pi$ decaying to charged lepton plus missing energy.
There are bounds from meson decays on such new couplings of order of $10^{-3}$ \cite{Farzan:2015doa,Pasquini:2015fjv,Lessa:2007up}.
In the future, more precise measurements of $K^-,\pi^- \to \ell_\alpha^- +{\rm missing~energy}$ can probe smaller values of this coupling, providing a way to test a significant part of the parameter space of our model.  The $Z^\prime$ boson can be produced and subsequently decay inside the supernova core \cite{Kamada:2015era}, affecting the radius of neutrinosphere and the duration of neutrino emission. Within present supernova uncertainties, this new interaction can be tolerated. Studying the detailed impact is beyond the scope of the present paper but we expect with improvements of theoretical and observational uncertainties the effect can be discerned providing another route to test the predictions of the model.
Lastly, a light $Z'$ coupled to neutrinos could have an impact on the astrophysical neutrino spectrum measured in IceCube, as emphasized in Refs.~\cite{Hooper:2007jr,Ioka:2014kca,Ng:2014pca,Araki:2014ona,DiFranzo:2015qea,Altmannshofer:2016brv}.

\section{Summary and outlook}
\label{summary}

Non-standard neutrino interactions have been widely discussed and constrained. Always luring in the shadows is the question of how to generate these interactions without violating the much stronger bounds from charged leptons. Here we have proposed the possibility that neutrinos mix with (at least) one new fermion which is coupled to a light mediator particle $Z'$. If this sub-GeV gauge boson further couples to one of the globally conserved charges of the SM, baryon number $B$, lepton number $L$ or $B-L$, we obtain the desired non-standard neutrino interactions, all the while suppressing effects in the charged-lepton sector.
Strong constraints from $\nu_{e,\mu}$--electron scattering experiments make it difficult -- but not impossible -- to obtain large NSI coefficients other than $\epsilon_{\tau\tau}$ if $Z'$ couples to lepton number.
 We have however found that the model in which $U(1)_B$ is gauged is quite  suitable  to obtain observable NSI $\epsilon_{\alpha\beta}^{u,d}$  for any flavor combination $\alpha,\beta$ as this model is not strongly constrained. In particular, the LMA-Dark solution can be realized by generating the single entry $\epsilon_{ee}\sim -1$. We have proposed a few ways to cancel the anomalies of $U(1)_B$.

For the minimal model with only one $U(1)^\prime$ charged Dirac fermion $\Psi$ mixed with active neutrinos, we predict $|\epsilon_{\alpha \beta}|=(\epsilon_{\alpha \alpha} \epsilon_{\beta \beta})^{1/2}$. Adding more such fermions, we still predict an inequality $|\epsilon_{\alpha \beta}|\leq (\epsilon_{\alpha \alpha} \epsilon_{\beta \beta})^{1/2}$. By proper choice of the sign of $g_Bg_\Psi$ and the phase of $y_\alpha y_\beta^*$, any  sign for diagonal elements and any phase for the off-diagonal elements of $\epsilon_{\alpha \beta}$ matrix can be obtained.

To mix neutrinos with the new Dirac fermion $\Psi$, we have to introduce a new electroweak doublet containing a charged scalar, $H^-$. At colliders with $s>4 M_{H^-}^2$, the $H^-H^+$ pair can be produced and subsequently decays to charged leptons plus $\Psi$, leading to a signal of an excess in  $l^- l^{+\prime} +{\rm missing~energy}$.
The discovery potential of such a scalar is explored in Ref.~\cite{Farzan:2010fw}.

The new couplings of $\nu$ with the light gauge boson $Z^\prime$ can have observable effects in precision measurements of charged meson decays to charged lepton plus missing energy. They can also affect the duration of neutrino emission at supernova explosions as well as Big Bang nucleosynthesis. Improvements in theoretical and experimental uncertainty in these observations can help to test a significant part of parameter space of the present model.
In fact, Ref.~\cite{Kamada:2015era} concludes $M_{Z^\prime}>\unit[5]{MeV}$ from the conservative cosmological  bound $\Delta N_\text{eff}>0.7$. We also discussed the bounds on neutrino--nucleon couplings from interaction rate of solar neutrinos in direct dark matter search experiments. We have found that although the present bound from CDMSlite is weak, future germanium or xenon based experiments such as SuperCDMS SNOLAB or LZ can help to probe a significant part of the parameter space of the present model which leads to sizeable $\epsilon_{\alpha \beta}$ for $M_{Z^\prime}>\unit[5]{MeV}$.

\subsection*{Acknowledgments}
The authors thank Ian Shoemaker for useful discussions in the early stages of this work.
This project has received funding from the European Union's Horizon 2020 research and innovation programme under the Marie Sk\l{}odowska-Curie grant agreement No.~674896 and No.~690575.
YF is also grateful to the ICTP associate office and Iran National Science Foundation (INSF) for partial financial support under contract 94/saad/43287.
JH is a postdoctoral researcher of the F.R.S.-FNRS.
We acknowledge the use of \texttt{Package-X}~\cite{Patel:2015tea} and \texttt{JaxoDraw}~\cite{Binosi:2003yf}.

\bibliographystyle{utcaps_mod}
\bibliography{references}
\end{document}